\documentclass[12pt]{iopart}
\usepackage{iopams}
\usepackage{graphicx}

\begin{document}
 
\title[CMB properties of the Hantzsche-Wendt Manifold]
{The Hantzsche-Wendt Manifold in Cosmic Topology}
\author{R.\ Aurich and S.\ Lustig}
 
\address{Institut f\"ur Theoretische Physik, Universit\"at Ulm,\\
Albert-Einstein-Allee 11, D-89069 Ulm, Germany}

\begin{abstract}
The Hantzsche-Wendt space is one of the 17 multiply connected spaces
of the three-dimensional Euclidean space $\mathbb{E}^3$.
It is a compact and orientable manifold which can serve as a model
for a spatial finite universe.
Since it possesses much fewer matched back-to-back circle pairs on the
cosmic microwave background (CMB) sky than the other compact flat spaces,
it can escape the detection by a search for matched circle pairs.
The suppression of temperature correlations $C(\vartheta)$
on large angular scales on the CMB sky is studied.
It is shown that the large-scale correlations are of the same order as
for the 3-torus topology but express a much larger variability.
The Hantzsche-Wendt manifold provides a topological possibility
with reduced large-angle correlations that can hide from searches
for matched back-to-back circle pairs.
\end{abstract}


\submitto{\CQG}

\section{Introduction}
\label{sec:intro}

An interesting aspect of cosmology concerns the global spatial
structure of our Universe, that is the question for its topology.
For review papers of cosmic topology and for discussions concerning
topological tests, see
\cite{Lachieze-Rey_Luminet_1995,Luminet_Roukema_1999,Levin_2002,%
Reboucas_Gomero_2004,Luminet_2008,Mota_Reboucas_Tavakol_2010,%
Mota_Reboucas_Tavakol_2011,Fujii_Yoshii_2011}.
Since the standard $\Lambda$CDM concordance model of cosmology is based on
the spatial flat space, the topological question might be restricted
to space forms admissible in the three-dimensional Euclidean space
$\mathbb{E}^3$.
There are 18 possible space forms which are denoted as $E_1$ to $E_{18}$
in \cite{Lachieze-Rey_Luminet_1995,Riazuelo_et_al_2004,Fujii_Yoshii_2011}.
The space $E_{18}$ is the usual simply connected Euclidean space
without compact directions.
The remaining 17 space forms possess compact directions and are thus
multiply connected.
The usual three-dimensional Euclidean space $\mathbb{E}^3$ can be considered
as their universal cover
which is tessellated by the multiply connected space forms into
spatial domains which have to be identified.
Eight of the 17 space forms are non-orientable manifolds which are usually
not taken into account in cosmology.
Thus there are 9 orientable multiply connected manifolds and 6 of them
are compact.
The focus is usually put on the 6 compact space forms $E_1$ to $E_6$.
From these the 3-torus topology $E_1$ has attracted the most attention
and its cosmological implications are well understood.
The $E_1$ space has the simplifying property that the statistical
cosmological properties are independent of the position of the
observer for which the statistics, e.\,g.\ of the cosmic microwave background
(CMB) radiation, is computed. 
This simplification is not possible for the other compact orientable
multiply connected manifolds $E_2$ to $E_6$ and the statistics of the
CMB simulations, on which the focus is put in this paper,
have to be computed for a sufficiently large number of observer positions
on the manifold in order to obtain a representative result
for such an inhomogeneous manifold.

A method to detect a non-trivial topology of our Universe is the
search for the circles-in-the-sky (CITS) signature
\cite{Cornish_Spergel_Starkman_1998b}.
Since the space is multiply connected,
the sphere from which the CMB radiation is emitted towards a given observer 
can overlap with another sphere belonging to a position in the
universal cover
which is, due to the topology, to be identified with that of the
considered observer.
The intersection of such spheres leads to circles on the CMB sky
where the temperature fluctuations are correlated according to the
assumed topology.
The simplest situation is realised by two circles whose centres are
antipodal on the CMB sky.
Such circle pairs are called ``back-to-back''.
This is the type of matched circle pairs which is the easiest one
to discover in CMB sky maps.
The non-back-to-back matched circle pairs have two further degrees
of freedom due to the position of the centre of the second circle.
This significantly enlarges the numerical effort in the CMB analysis and,
in addition, increases the background of accidental correlations
which can swamp the true signal of a matched circle pair.
There are a lot of papers devoted to the CITS search
\cite{
Roukema_1999,Roukema_2000,
Cornish_Spergel_Starkman_Komatsu_2003,
Roukema_et_al_2004,
Aurich_Lustig_Steiner_2005b,
Then_2006a,
Key_Cornish_Spergel_Starkman_2007,
Aurich_Janzer_Lustig_Steiner_2007,
Bielewicz_Banday_2011,
Vaudrevange_Starkman_Cornish_Spergel_2012,
Rathaus_BenDavid_Itzhaki_2013,
Aurich_Lustig_2013,
Planck_Topo_2013}.
The result is that there is no convincing hint for a matched
back-to-back circle pair in the CMB data
with a radius above $25^\circ\dots 30^\circ$.
Smaller back-to-back circle pairs are not detectable
\cite{Aurich_Lustig_2013}.
This leads to the question whether space forms without
back-to-back circle pairs can get missed by these searches.
The non-back-to-back circle search in
\cite{Vaudrevange_Starkman_Cornish_Spergel_2012}
does not find hints of such a topology.
However, it is shown in \cite{Aurich_Lustig_2013} that the search grid
in \cite{Vaudrevange_Starkman_Cornish_Spergel_2012} is too coarse
such that even large back-to-back circles with radii up to $50^\circ$
are not found (see figure 4 in \cite{Aurich_Lustig_2013}).
The analysis in \cite{Aurich_Lustig_2013} refers to back-to-back circle pairs,
but the increased background in the general case worsen the detectability
of non-back-to-back circles.
Furthermore, due to foreground contamination in two regions of the
CMB map, the search in \cite{Vaudrevange_Starkman_Cornish_Spergel_2012}
excludes circles which intersect these regions.
Thus it is safe to say that \cite{Vaudrevange_Starkman_Cornish_Spergel_2012}
does not find a hint in favour of such a topology,
but it cannot exclude one.

Restricting to the Euclidean case with its 6 compact orientable space forms, 
one can ask which topology does not possess back-to-back circle pairs
with radii above $25^\circ\dots 30^\circ$.
The answer depends on the value of the parameters $L_i$,
which define the size and shape of the Dirichlet cell,
see equation (\ref{Eq:Def_Gamma}) below.
Let us assume that these parameters are chosen in such a way
that all spatial dimensions of the Dirichlet cell are of the same order.
This ensures that the largest circle pairs arise from the identification
of the faces of the Dirichlet cell.
The topologies $E_1$ to $E_5$ identify at least two pairs of faces
by pure translations, i.\,e.\ without an accompanying rotation.
These translations lead to back-to-back circle pairs for all observer positions
even in the case of an inhomogeneous manifold.
For example, the spaces $E_4$ and $E_5$, belonging to a hexagonal tiling of
the Euclidean space, possess three back-to-back circle pairs due to the
three pairs of faces that are identified by pure translations. 
In this respect, the manifold $E_6$ is special since every identification
of a pair of faces is defined by a translation and a rotation by $\pi$,
i.\,e.\ a so-called half-turn corkscrew motion.
The space form $E_6$ is also called Hantzsche-Wendt space
\cite{Hantzsche_Wendt_1935} and is the topic of this paper.

Therefore, the Hantzsche-Wendt space is a candidate for cosmic topology
worth studying its implications on the CMB sky.
The CMB angular power spectrum is computed for a single observer position 
in \cite{Scannapieco_Levin_Silk_1999} for low multipoles,
and a suppression of temperature correlations on large angular scales
is found.
Besides this, there are no further CMB analyses in literature,
although \cite{Riazuelo_et_al_2004} describes the eigenmodes of that
space form allowing the computation of CMB fluctuations for an observer
at the origin of the coordinate system.
The aim of this paper is to provide a CMB analysis
of the temperature correlations on large angular scales
for a huge sample of observer positions in order to allow
a comparison with the CMB observations.

\section{Hantzsche-Wendt Topology}
\label{Hantzsche-Wendt_Topology}

\begin{figure}
\begin{center}
\vspace*{0pt}
\begin{minipage}{15.8cm}
\hspace*{0pt}\includegraphics[width=15.8cm]{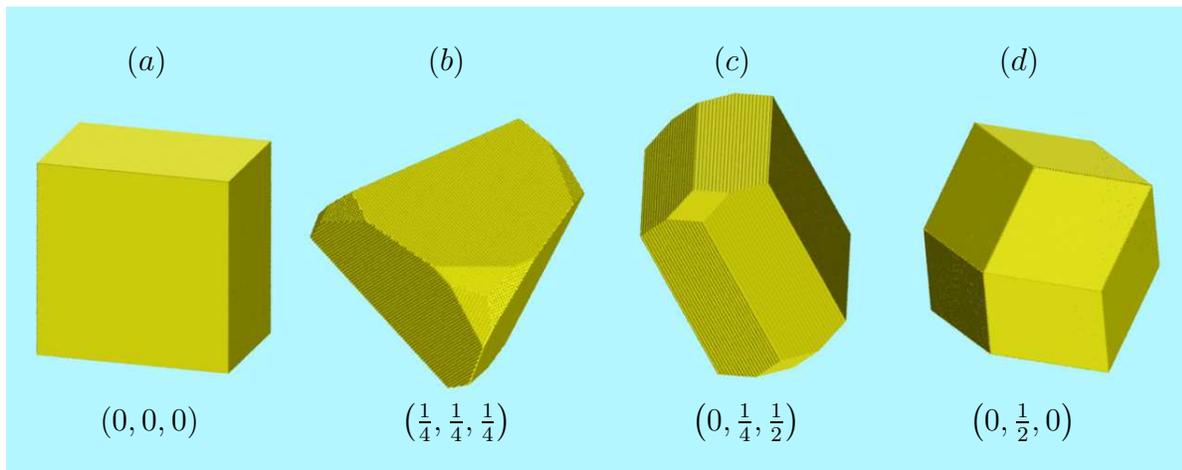}
\end{minipage}
\put(-404,68){$(a)$}
\put(-414,-68){$\left(0,0,0\right)$}
\put(-290,68){$(b)$}
\put(-300,-68){$\left(\frac 14,\frac 14,\frac 14\right)$}
\put(-182,68){$(c)$}
\put(-190,-68){$\left(0,\frac 14,\frac 12\right)$}
\put(-74,68){$(d)$}
\put(-85,-68){$\left(0,\frac 12,0\right)$}
\end{center}
\vspace*{0pt}
\caption{\label{Fig:HW_Dirichlet}
The Dirichlet domain is shown for four different observer positions,
i.\,e.\ different positions on the manifold taken as the origin
of the reference system.
The topological parameters are set equal, i.\,e.\ $L_x=L_y=L_z$.
The positions are indicated in the figure by the relative positions $(x,y,z)$.
While for $(x,y,z)=(0,0,0)$ the Dirichlet domain is a cuboid,
a rhombic dodecahedron is obtained for $(x,y,z)=\left(0,\frac 12,0\right)$.
}
\end{figure}

The Hantzsche-Wendt manifold $\mathbb{E}^3/\Gamma$ is the quotient of the
Euclidean space $\mathbb{E}^3$ by the holonomy group $\Gamma$
which is generated by the transformations
\begin{eqnarray}\nonumber 
(\tilde x,\tilde y,\tilde z) & \rightarrow & (\tilde x+L_x,-\tilde y,-\tilde z)
\\ \label{Eq:Def_Gamma}
(\tilde x,\tilde y,\tilde z) & \rightarrow &
(-\tilde x,\tilde y+L_y,-(\tilde z+L_z))
\\ \nonumber 
(\tilde x,\tilde y,\tilde z) & \rightarrow &
(-(\tilde x+L_x),-(\tilde y+L_y),\tilde z+L_z)
\end{eqnarray}
with the topological scale  defining parameters $L_x$, $L_y$, and $L_z$.
These generators are half-turn corkscrew motions,
since each transformation contains a rotation by $\pi$,
and lead to a fixed point free symmetry group $\Gamma$.
The definition (\ref{Eq:Def_Gamma}) of the generators of $\Gamma$
agrees with that in
\cite{Levin_Scannapieco_Silk_1998,Scannapieco_Levin_Silk_1999} but,
however, differs from that in \cite{Riazuelo_et_al_2004,Fujii_Yoshii_2011}.
The various definitions of the generators of $\Gamma$ all lead to the
same manifold, of course, but correspond to different positions on the manifold
which are taken as the origin of the reference system.
It is convenient to describe the positions $\vec x$ by dimensionless
coordinates $x$, $y$, $z\in [-0.5,0.5]$, such that a position $(x,y,z)$
with $\vec x := (x L_x, y L_y, z L_z)$ refers to corresponding
positions in Hantzsche-Wendt spaces defined by different sets of
topological length scales.
The Dirichlet domain with respect to a position $\vec x_o$ is defined as
the set of points $\vec x$ such that
$d(\vec x_o,\vec x) \leq d(\vec x_o,\gamma \vec x) $ $\forall \gamma\in\Gamma$.
Figure \ref{Fig:HW_Dirichlet} shows four examples of the Dirichlet domain
based on the generators (\ref{Eq:Def_Gamma}).
The geometrical shape of the Dirichlet domain possesses only for special chosen
positions $(x,y,z)$ highly symmetrical domains, e.\,g.\
a cuboid for $(0,0,0)$ or a rhombic dodecahedron for $(0,\frac 12,0)$
for $L_x=L_y=L_z$.

\begin{figure}
\begin{center}
\vspace*{0pt}
\begin{minipage}{15.8cm}
\hspace*{0pt}\includegraphics[width=15.8cm]{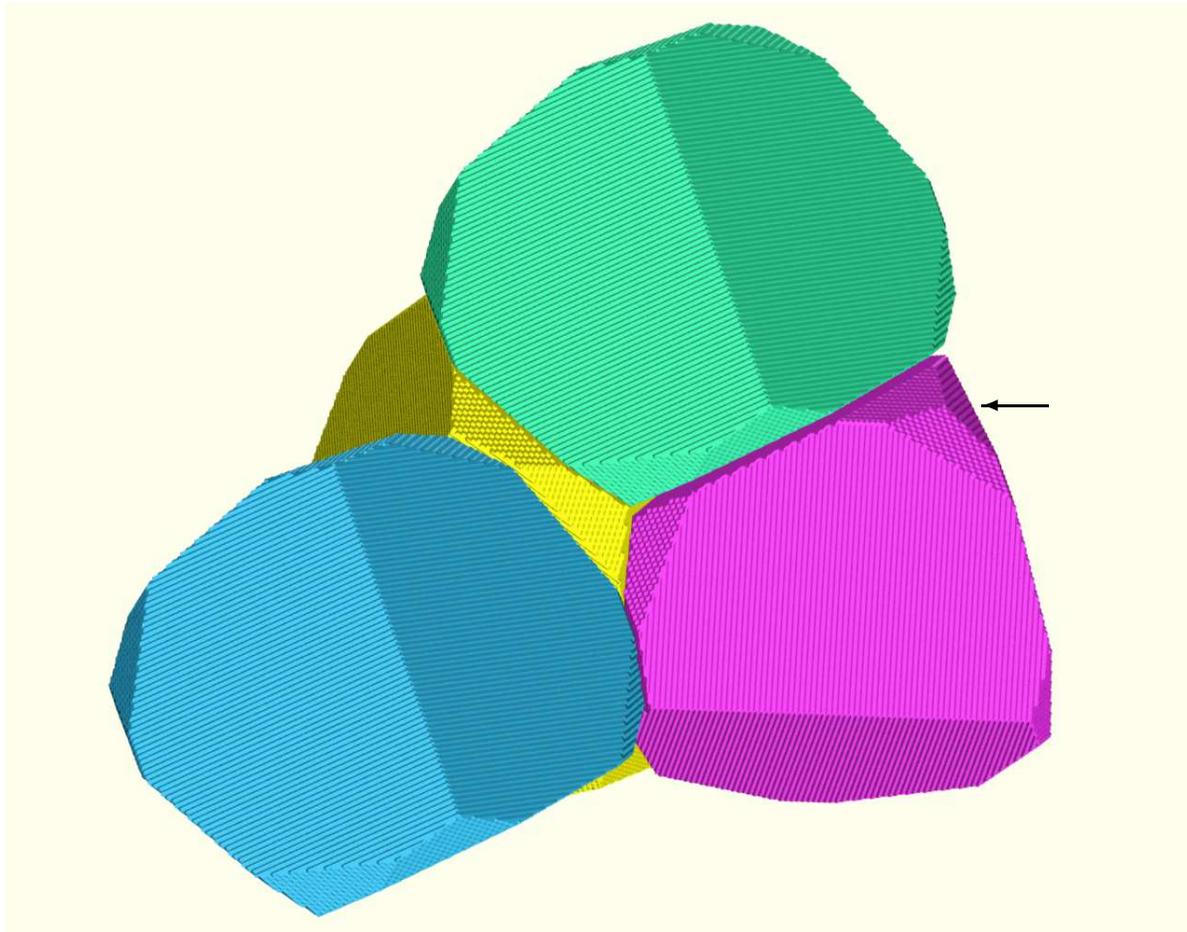}
\put(-55,200){\thicklines \vector(-1,0){25}}
\end{minipage}
\end{center}
\vspace*{0pt}
\caption{\label{Fig:HW_Dirichlet_tess}
The Dirichlet domain is shown for the observer position
$(x,y,z)=\left(\frac 14,\frac 14,\frac 14\right)$
with three adjacent domains.
The arrow points towards a corner on the Dirichlet domain,
where three small patches meet.
This corner fits into the region shown in the centre of the image,
where the yellow, green and purple domains are in direct contact.
}
\end{figure}

At a first sight, one might get the impression that the domains
shown in figure \ref{Fig:HW_Dirichlet}(b) or \ref{Fig:HW_Dirichlet}(c)
might not tessellate the Euclidean space.
To demonstrate that this is indeed the case,
figure \ref{Fig:HW_Dirichlet_tess} shows for the case
$(x,y,z)=\left(\frac 14,\frac 14,\frac 14\right)$,
i.\,e.\ figure \ref{Fig:HW_Dirichlet}(b),
how four adjacent Dirichlet domains stick together according to
the group $\Gamma$ generated by (\ref{Eq:Def_Gamma}).
One interesting corner of the Dirichlet domain is indicated by the arrow
in figure \ref{Fig:HW_Dirichlet_tess},
which is formed by three small surface patches.
This surface structure matches the joint of the three Dirichlet domains
shown in the centre of figure \ref{Fig:HW_Dirichlet_tess}.
In this way, the next Dirichlet domain would exactly match to this dip,
so that the Euclidean space is tessellated with no overlaps and no gaps.

\begin{figure}
\begin{center}
\vspace*{-15pt}
\begin{minipage}{18cm}
\hspace*{-30pt}\includegraphics[width=10.0cm]{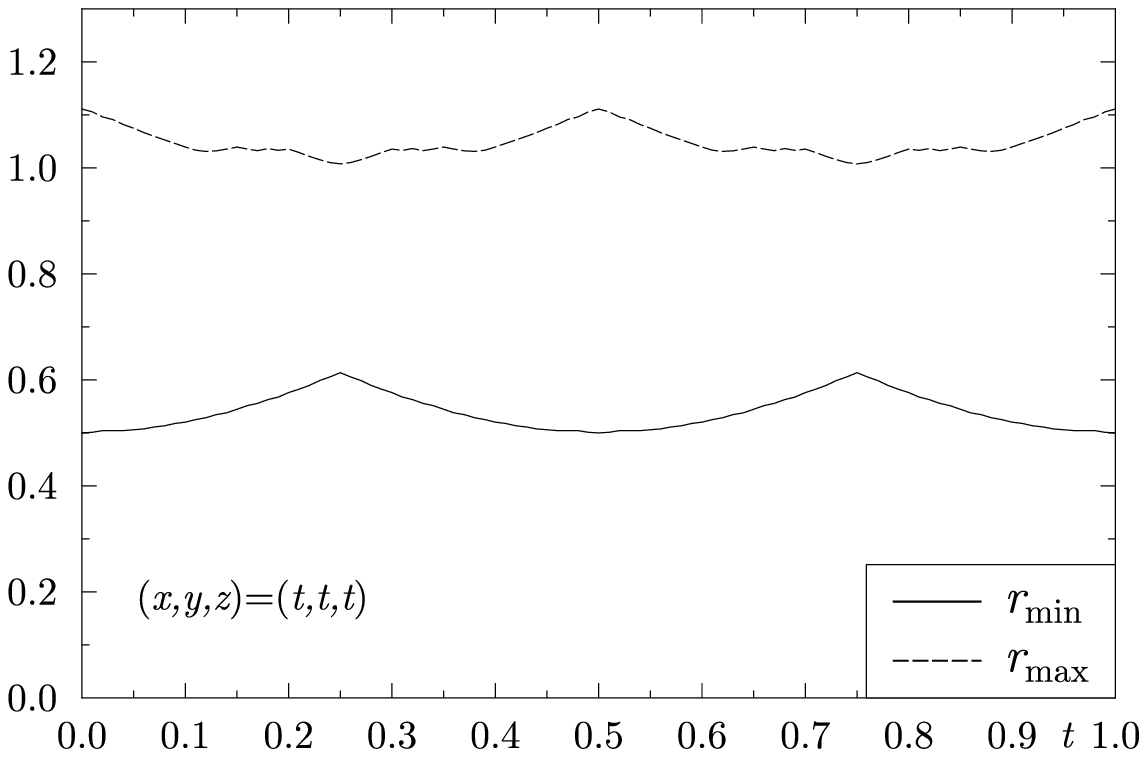}
\hspace*{-55pt}\includegraphics[width=10.0cm]{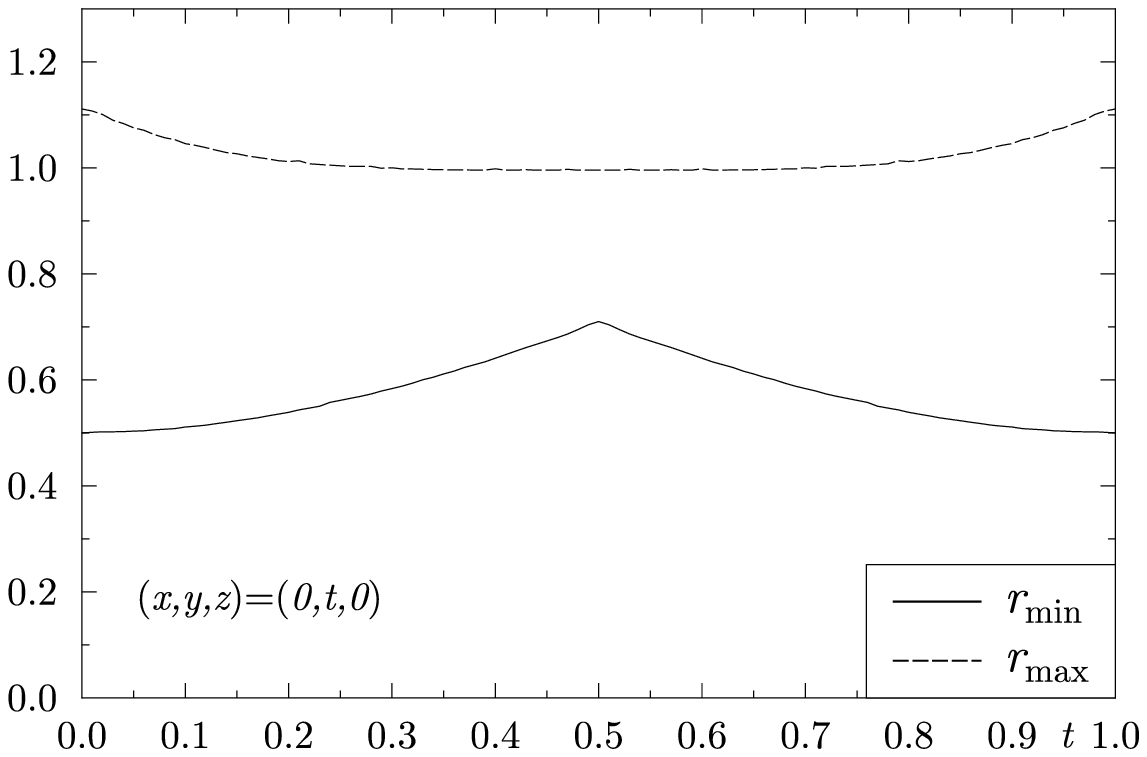}
\end{minipage}
\put(-480,50){$(a)$}
\put(-246,50){$(b)$}
\end{center}
\vspace*{-20pt}
\caption{\label{Fig:radius_min_max}
The distance to the surface of the Dirichlet domain measured from
the observer position is shown.
The minimal distance is plotted as $r_{\hbox{\scriptsize min}}$ and the
maximal distance as $r_{\hbox{\scriptsize max}}$.
The position of the observer is varied along the line
$(x,y,z)=(t,t,t)$ in panel (a) and along $(x,y,z)=(0,t,0)$ in panel (b)
for $t\in[0,1]$.
The topological parameters are set as $L_x=L_y=L_z=1$.
}
\end{figure}

Another possibility to obtain information about the shape of the
Dirichlet domain is to consider the range of distances $r$ of
the observer positions from the surface of the Dirichlet domain.
In figure \ref{Fig:radius_min_max} the values of the minimum of
these distances $r_{\hbox{\scriptsize min}}$ and of the
maximum $r_{\hbox{\scriptsize max}}$ are plotted for two sets of
observer positions.
Panel \ref{Fig:radius_min_max}(a) shows the distances for the
line parameterised as $(x,y,z)=(t,t,t)$, $t\in[0,1]$,
which for $t=0$ and $t=\frac 14$ agrees with the observer positions
of figures \ref{Fig:HW_Dirichlet}(a) and \ref{Fig:HW_Dirichlet}(b).
One observes that for $t=0$ the width of the distribution of the
distances is maximal that is $r_{\hbox{\scriptsize min}}$ is minimal
and $r_{\hbox{\scriptsize max}}$ is maximal.
The reverse situation occurs for $t=\frac 14$ where the width is minimal.
Figure \ref{Fig:radius_min_max}(b) shows the  distances for the line
$(x,y,z)=(0,t,0)$, where the example for $t=\frac 12$ is displayed
in figure \ref{Fig:HW_Dirichlet}(d).
This position belongs to the rhombic dodecahedron having a minimal
width $r_{\hbox{\scriptsize max}}-r_{\hbox{\scriptsize min}}$
as figure \ref{Fig:radius_min_max}(b) reveals.

In the following, we will measure $L_x$, $L_y$, and $L_z$ in units of
the Hubble length $L_{\hbox{\scriptsize H}} = c/H_0$.
The parameters of the $\Lambda$CDM concordance model used in this paper
are obtained from  Table 8, column ``WMAP+BAO+$H_0$'' of
\cite{Jarosik_et_al_2010} stating the values
$\Omega_{\hbox{\scriptsize bar}} = 0.0456$,
$\Omega_{\hbox{\scriptsize cdm}} = 0.227$, 
a present day reduced Hubble constant $h=0.704$,
a reionization optical depth $\tau=0.087$,
and a scalar spectral index $n_s=0.963$.
These parameters lead to a diameter of the surface of last scattering of
$D_{\hbox{\scriptsize sls}} = 6.605$ in units of
$L_{\hbox{\scriptsize H}} \simeq 4.258\hbox{ Gpc}$.
The order of the topological length scale should not significantly be
above $D_{\hbox{\scriptsize sls}}$ in order to leave a signature on the CMB sky
\cite{Fabre_Prunet_Uzan_2013}.
Although the cosmological parameters of $\Lambda$CDM concordance model
differ for the different WMAP data releases and also depend on the
inclusions of other cosmological observations, it is not expected
that the topological results referring to the largest scales are
affected by these changes.
This statement also applies to the parameter set given by the
Planck collaboration \cite{Planck_Cosmo_Parameters_2013}.

The fact that all three generators are half-turn corkscrew motions
leads to the important consequence that the matched circles pairs
with the largest radii are not of the back-to-back type.
The difficulties in detecting such matched circles pairs motivate
the CMB analysis of the Hantzsche-Wendt space as discussed in the
Introduction.
Let us now discuss how many matched back-to-back circle pairs exist
for a given Hantzsche-Wendt manifold on the corresponding CMB sky.
Although, for general observer positions,
the generators do not lead to back-to-back circles,
it is clear that the group element obtained by applying one of the
generators twice, will lead to a back-to-back circle pair
since the two rotations by $\pi$ lead to a full turn.
That transformation, however, has twice the length scale $L_{x,y,z}$
and matched circle pairs with smaller radii in turn.
This example shows that the group $\Gamma$ generated by (\ref{Eq:Def_Gamma})
contains pure translations, i.\,e.\ without a corkscrew motion.
The point is, however, that these group elements lead to circles
with smaller radii.

\begin{figure}
\begin{center}
\vspace*{-15pt}
\begin{minipage}{18cm}
\hspace*{-30pt}\includegraphics[width=10.0cm]{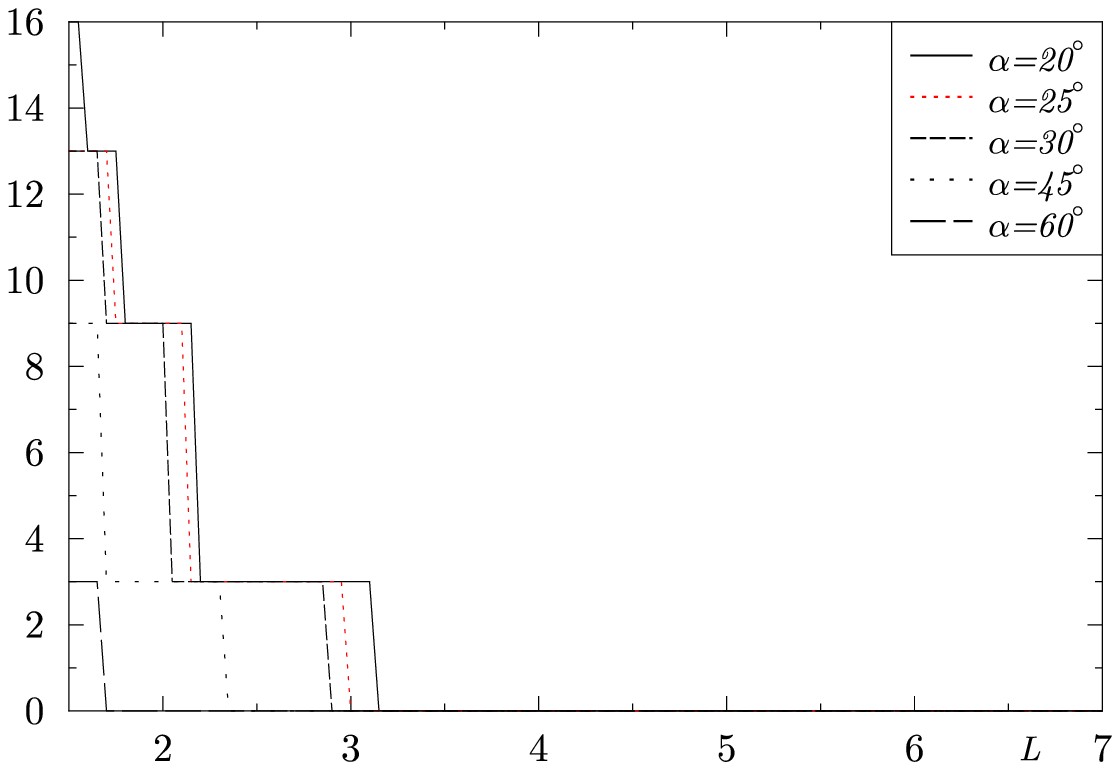}
\hspace*{-55pt}\includegraphics[width=10.0cm]{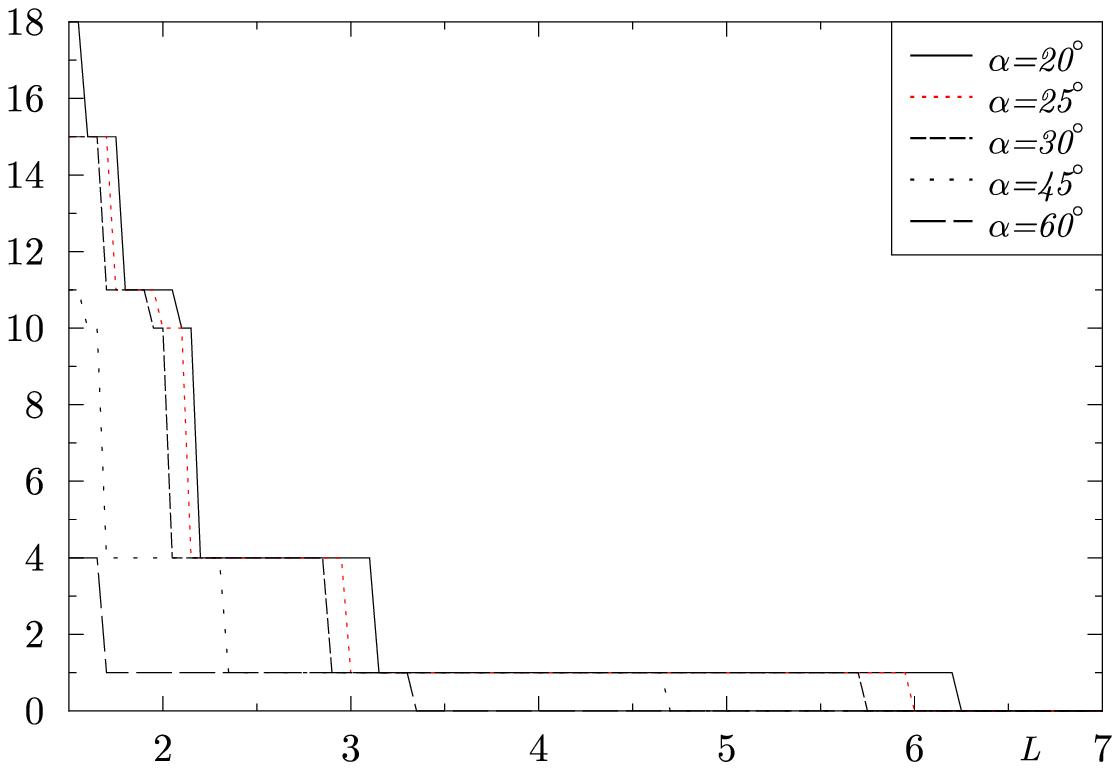}
\end{minipage}
\put(-450,50){$(a)$}
\put(-202,50){$(b)$}
\end{center}
\vspace*{-20pt}
\caption{\label{Fig:cits_back_to_back}
The number $N_\alpha$ of matched back-to-back circle pairs with a radius
greater than $\alpha$ is plotted for the Hantzsche-Wendt space
and several values of $\alpha$.
Since this multiply connected space is inhomogeneous,
the number $N_\alpha$ is computed for a large sample of positions on
the Hantzsche-Wendt manifold defined by the length scale $L$.
Panel (a) shows the minimum of $N_\alpha$ taken over all positions with
fixed $L$, whereas panel (b) shows the maximum of $N_\alpha$.
}
\end{figure}

\begin{figure}
\begin{center}
\vspace*{-15pt}
\begin{minipage}{10cm}
\includegraphics[width=10.0cm]{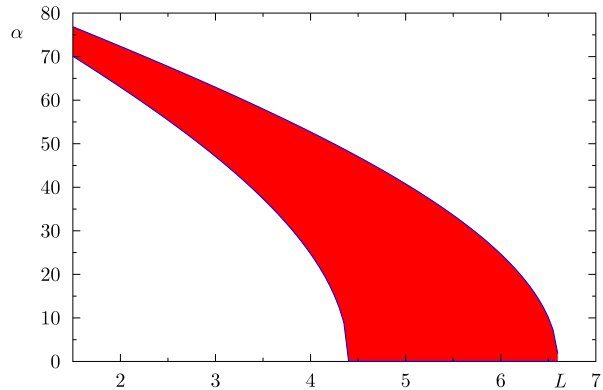}
\end{minipage}
\end{center}
\vspace*{-20pt}
\caption{\label{Fig:cits_non_back_to_back}
The range of radii of the largest non-back-to-back circle pair
is plotted in dependence on the length scale $L$.
The distribution is obtained from the same mesh of observer positions
as in figure \ref{Fig:cits_back_to_back}.
For $L>4.4$ there are positions where the radius of the largest
non-back-to-back circle pair approaches zero,
so that no non-back-to-back circle pair occurs.
}
\end{figure}

Let us define $N_\alpha$ as the number of matched back-to-back circle pairs
with a radius greater than $\alpha$.
The larger the value of $N_\alpha$ for $\alpha > 25^\circ\dots 30^\circ$ is,
the more increases the likelihood that the topology will be detected
by a CITS search.
The dependence of $N_\alpha$ on the size of the Hantzsche-Wendt space
is shown in figure \ref{Fig:cits_back_to_back}.
There the symmetrical case $L=L_x=L_y=L_z$ is considered, and
the variable $L$ is used as the size parameter.
One could compare its value with the diameter of the surface of last scattering
which is $D_{\hbox{\scriptsize sls}} = 6.605$.
Since the Hantzsche-Wendt space is inhomogeneous,
it does not suffice to compute the number $N_\alpha$ for one observer position.
One obtains different values for $N_\alpha$ for different positions.
A cubic mesh with $51^3$ positions is generated covering the
Hantzsche-Wendt space,
and the values of $N_\alpha$ are calculated thereon.
Panel (a) of figure \ref{Fig:cits_back_to_back} shows the smallest
number of $N_\alpha$ that occurs on the grid for various circle radii $\alpha$
as a function of $L$.
The other panel shows the corresponding maximal values.
One observes in panel (a) that there is no back-to-back circle pair
for $L>3$ with a radius $\alpha > 25^\circ$,
at least for some observer positions,
while there is at most one such circle pair, see panel (b).
This panel shows that one back-to-back circle pair exists for $L$
up to $L\simeq 6$ for special observer positions.
This can be read off from the definition of the group
generators (\ref{Eq:Def_Gamma}).
Consider the first generator in (\ref{Eq:Def_Gamma}) and set $y=z=0$
such that the selected position is on the axis of rotation belonging
to this generator.
Then the positions on the axis of rotation, i.\,e.\ the $x$ axis,
have all to be identified if their distances are multiples of $L_x$
from each other.
This leads to back-to-back circles with circle centres on the axis of rotation
for such special selected observer positions.
But an arbitrarily chosen observer will not observe back-to-back circle pairs
for $L>3$.
The radius of the largest non-back-to-back circle pair also depends on the
observer position.
Figure \ref{Fig:cits_non_back_to_back} shows the range of radii found on the
same set of observer positions that is used in the computation
of figure \ref{Fig:cits_back_to_back}.
Above $L=4.4$ there are positions without a non-back-to-back circle pair
as can be received from figure \ref{Fig:cits_non_back_to_back} showing
that the radius of the largest pair can approach zero.
The angle of antipodicity covers the large interval from almost zero
up to $96.37^\circ$ depending on the position, of course.
Other considerations concerning the largest circle pair can be found in
\cite{Mota_Reboucas_Tavakol_2010,Mota_Reboucas_Tavakol_2011}.

One thus concludes that the Hantzsche-Wendt space with $L>3$ will not be
discovered by searches for matched back-to-back circle pairs.
Even for slightly smaller topological length scales,
the matched back-to-back circle pairs possess small radii
so that large parts of a circle could be obscured by residual foregrounds.
The original motivation for studying cosmic topology derives mainly
from the natural ability of multiply connected spaces to suppress the
temperature correlations on large angular scales on the CMB sky.
At least, if the multiply connected spaces are smaller than the
surface of last scattering.
It is therefore important to compute these CMB temperature correlations
for the Hantzsche-Wendt manifold for various values of $L$ and
for a large set of observer positions in order to allow a representative
comparison with the CMB observations.
This requires the computation of the eigenmodes of the Laplacian
in the Hantzsche-Wendt space which is the topic of the next section.

\section{Eigenmodes in the Hantzsche-Wendt Space}

In Euclidean space the eigenmodes can be represented as
linear combinations of plane waves.
They are discussed in
\cite{Levin_Scannapieco_Silk_1998,Scannapieco_Levin_Silk_1999,%
Riazuelo_et_al_2004,Riazuelo_Uzan_Lehoucq_Weeks_2004},
where they are given with respect to the origin of the
considered coordinate system.
For a CMB analysis build on a large set of observer positions,
it is, however, necessary to effectively calculate the eigenmodes
in the spherical basis
with respect to an arbitrarily chosen observer position.
This can be achieved in the following way.

In the Hantzsche-Wendt space the eigenmodes can be expressed
as a linear combination of four plane waves
\begin{equation}
\label{Eq:plane_wave}
\Psi_{\vec k}(\vec x\,) \; = \; \gamma
\sum_{j=1}^4 \exp(\hbox{i}\, \vec k_j \vec x\,) \, \exp(\hbox{i}\,\phi_j)
\hspace{10pt} .
\end{equation}
Since the eigenmodes have to be invariant under the action of the
holonomy group $\Gamma$ generated by (\ref{Eq:Def_Gamma}),
the wave numbers $\vec k$ take on the values
\begin{equation}
\label{Eq:wave_number}
\vec k \; = \; (k_x,k_y,k_z) \; = \;
\pi \, \left( \frac{n_x}{L_x}, \frac{n_y}{L_y}, \frac{n_z}{L_z} \right)
\hspace{10pt} \hbox{ with } \hspace{10pt}
(n_x,n_y,n_z) \in {\mathbb{Z}^3}
\end{equation}
with some restrictions to $(n_x,n_y,n_z)$.
The factor $\gamma$ in (\ref{Eq:plane_wave}) takes on the value
$1/\sqrt 2$ if two of the numbers $n_x,n_y,n_z$ are equal zero
and is one otherwise.
This definition of $\gamma$ leads to normalised eigenmodes.
The allowed values of the numbers $n_x,n_y,n_z$ are
$n_x,n_y,n_z\in {\mathbb N}$, $-n_x,-n_y,-n_z\in {\mathbb N}$,
and $n_x=0$, $n_y,n_z\in {\mathbb N}$ and cyclic permutations thereof.
Finally, the last possibility is $n_x=n_y=0$, $n_z\in 2{\mathbb N}$ together
with cyclic permutations.
The wave numbers $\vec k_j$ and the phases $\phi_j$ occurring in
the plane wave representation (\ref{Eq:plane_wave}) are
related to $\vec k$, equation (\ref{Eq:wave_number}), by
\begin{eqnarray}
\nonumber
\vec k_1 & = & (k_x,k_y,k_z)
\hspace{29pt} , \hspace{10pt}
\phi_1 \; = \; 0
\\ \nonumber
\vec k_2 & = & (k_x,-k_y,-k_z)
\hspace{10pt} , \hspace{10pt}
\phi_2 \; = \; \pi\, n_x
\\ \nonumber
\vec k_3 & = & (-k_x,k_y,-k_z)
\hspace{10pt} , \hspace{10pt}
\phi_3 \; = \; \pi (n_y+n_z)
\\ \nonumber
\vec k_4 & = & (-k_x,-k_y,k_z)
\hspace{10pt} , \hspace{10pt}
\phi_4 \; = \; \pi (n_x+n_y+n_z)
\hspace{10pt} .
\end{eqnarray}
Let us expand in the plane wave for a general observer position $\vec x_o$,
i.\,e.\
\begin{equation}
\label{Eq:plane_wave_shifted}
\Psi_{\vec k}(\vec x\,) \; = \;
\gamma \sum_{j=1}^4 \exp(\hbox{i}\, \vec k_j \vec x_o) \,
\exp\big(\hbox{i}\, \vec k_j(\vec x-\vec x_o)\big) \, \exp(\hbox{i}\,\phi_j)
\hspace{10pt} ,
\end{equation}
the factor $\exp(\hbox{i}\, \vec k_j (\vec x-\vec x_o))$
with respect to the spherical basis using
\begin{equation}
\label{Eq:plane_wave_spherical}
\exp(\hbox{i}\, \vec k (\vec x-\vec x_o)) \, \; = \;
4\pi \sum_{l=0}^\infty \sum_{m=-l}^l \hbox{i}^l \, j_l(kr) \,
Y_{lm}^\star(\hat{n}_k) \, Y_{lm}(\hat{n}_r)
\hspace{10pt} ,
\end{equation}
with $r =|\vec x-\vec x_o|$ and $k = |\vec k|$.
Furthermore, $j_l$ denotes the spherical Bessel function, $Y_{lm}(\hat{n})$
the spherical harmonics, and the unit vector $\hat{n}$ provides
the usual angles $\vartheta$ and $\phi$ as arguments for $Y_{lm}$.

Note that the wave numbers $\vec k_2$, $\vec k_3$, and $\vec k_4$
are obtained from $\vec k_1$ by a half-turn rotation around the
$x$-, $y$-, and $z$-axis in the $k$ space, respectively,
\begin{equation}
\label{Eq:k_rotation}
\vec k_2 \; = \; {\cal R}_x(\pi) \, \vec k_1
\hspace{10pt} , \hspace{10pt}
\vec k_3 \; = \; {\cal R}_y(\pi) \, \vec k_1
\hspace{10pt} \hbox{ and } \hspace{10pt}
\vec k_4 \; = \; {\cal R}_z(\pi) \, \vec k_1
\hspace{10pt} .
\end{equation}
This allows to rewrite $Y_{lm}(\hat{n}_j)$ with
$\hat{n}_j:= \vec k_j/|\vec k_j|$ as
\begin{equation}
\label{Eq:Wigner_rotation}
Y_{lm}(\hat{n}_j) \; = \;
\sum_{m'=-l}^l Y_{lm'}(\hat{n}_1) \, D_{mm'}^l(\alpha_j,\beta_j,\gamma_j)
\hspace{10pt} , \hspace{10pt}
j=2,3,4
\hspace{10pt} ,
\end{equation}
where
$D_{mm'}^l(\alpha,\beta,\gamma)=
e^{-\hbox{\scriptsize i}m'\alpha} d_{mm'}^l(\beta) e^{-\hbox{\scriptsize i}m\gamma}$
denote the Wigner polynomials.
By setting the angles $\alpha_j,\beta_j,\gamma_j$ according to the rotations
(\ref{Eq:k_rotation}),
the plane wave can be written as
\begin{eqnarray}
\label{Eq:PlaneWaveComplex}
\Psi_{\vec k}(r, \hat n_r,\vec x_o) & = &
4\pi \gamma \sum_{l=0}^\infty \sum_{m=-l}^l \hbox{i}^l \, j_l(kr) \,
Y_{lm}(\hat{n}_r) \Big\{
e^{\hbox{\scriptsize i} \vec k_1 \vec x_o} Y_{lm}^\star(\hat{n}_1) \, 
\\ \nonumber & & \hspace{10pt}
+ \;
(-1)^{l+m} \, e^{\hbox{\scriptsize i} (\vec k_2 \vec x_o+\phi_2)} Y_{lm}(\hat{n}_1)
 \;+ \;
(-1)^l \, e^{\hbox{\scriptsize i} (\vec k_3 \vec x_o+\phi_3)} Y_{lm}(\hat{n}_1)
\\ \nonumber & & \hspace{50pt}
+ \;
(-1)^{m} \, e^{\hbox{\scriptsize i} (\vec k_4 \vec x_o+\phi_4)} Y_{lm}^\star(\hat{n}_1)
\Big\}
\hspace{10pt} .
\end{eqnarray}
The next step is to construct real eigenmodes
from the expression (\ref{Eq:PlaneWaveComplex}).
One defines a complex coefficient $c$ and obtains
the real eigenmode
\begin{equation}
\label{Eq:plane_wave_real}
\Psi^R_{\vec k}(r, \hat n_r,\vec x_o) \; := \;
c \, \Psi_{\vec k}(r, \hat n_r,\vec x_o) \, + \,
c^\star\, \Psi_{-\vec k}(r, \hat n_r,\vec x_o)
\hspace{10pt} .
\end{equation}
With the observer position dependent phases
\begin{equation}
\label{Eq:coefficient_alpha}
\alpha \; := \;
e^{\hbox{\scriptsize i} \vec k_1 \vec x_o} \, + \,
(-1)^m e^{\hbox{\scriptsize i}(\vec k_4 \vec x_o+\phi_4)}
\end{equation}
and
\begin{equation}
\label{Eq:coefficient_beta}
\beta \; := \;
(-1)^m e^{\hbox{\scriptsize i}(\vec k_2 \vec x_o+\phi_2)}  \, + \,
e^{\hbox{\scriptsize i}(\vec k_3 \vec x_o+\phi_3)}
\hspace{10pt} ,
\end{equation}
the real eigenmode can be written as
\begin{eqnarray} \nonumber
\Psi^R_{\vec k}(r, \hat n_r,\vec x_o) & = &
4\pi\gamma \sum_{l=0}^\infty \sum_{m=-l}^l \hbox{i}^l\, Y_{lm}(\hat{n}_r) \,
j_l(kr) \; \Big\{
Y_{lm}^\star(\hat{n}_1) \, \left[ c \alpha + (-1)^l c^\star \alpha^\star \right]
\\ & & \hspace{90pt} + \;
\label{Eq:plane_wave_real_spherical}
Y_{lm}(\hat{n}_1) \, \left[ (-1)^l c \beta + c^\star \beta^\star \right] \Big\}
\hspace{10pt} .
\end{eqnarray}
After defining the spherical expansion coefficients
\begin{equation}
\label{Eq:a_lm} \hspace*{-40pt}
\xi_{lm}^{\vec k}(\vec x_o) \; := \;
\hbox{i}^l \gamma \Big\{
Y_{lm}^\star(\hat{n}_1) \, \left[ c \alpha + (-1)^l c^\star \alpha^\star \right]
\; + \;
Y_{lm}(\hat{n}_1) \, \left[ (-1)^l c \beta + c^\star \beta^\star \right] \Big\}
\hspace{10pt} ,
\end{equation}
one can write the eigenmode as
$\Psi^R_{\vec k}(r, \hat n_r,\vec x_o) =
\sum_{l=0}^\infty \sum_{m=-l}^l \xi_{lm}^{\vec k}(\vec x_o) \, R_{kl}(r) \,
Y_{lm}(\hat{n}_r)$
with the radial function $R_{kl}(r) = 4\pi j_l(kr)$. 

In CMB simulations a Gaussian random superposition of the eigenmodes
is required.
One has to ensure that all eigenmodes contribute with the same statistical
weight.
This can be realised by Gaussian random coefficients
$c = (c_R + \hbox{i} c_I)/\sqrt 2$, such that real and imaginary parts
$c_R$ and $c_I$ are Gaussian random variables with zero mean and unit variance
and thus $\left<|c|^2\right> = 1$.
However, one special case has to be taken into account.
If at least one of the components $k_x$, $k_y$ or $k_z$
of the wave number $\vec k$ vanishes,
the function in (\ref{Eq:PlaneWaveComplex}) is already real
or purely imaginary.
In that case the coefficient $c$ has to be multiplied by an additional factor
$1/\sqrt 2$ to ensure the same statistical weight.

\section{CMB Statistics}

The CMB anisotropies $\delta T(\hat n_r)$ are expanded with respect to
the spherical harmonics
\begin{equation}
\label{Eq:CMB_expansion}
\delta T(\hat{n}_r) \; = \;
\sum_{l=0}^\infty \sum_{m=-l}^l
a_{lm} \, Y_{lm}(\hat{n}_r)
\end{equation}
with
\begin{equation}
\label{Eq:alm}
a_{lm} \; = \; \sum_{\vec k} T_l(k)\,\sqrt{P(k)} \, \xi_{lm}^{\vec k}(\vec x_o)
\hspace{10pt} ,
\end{equation}
where the Gaussian random coefficients are contained in $\xi_{lm}^{\vec k}(\vec x_o)$.
Here, $P(k)\sim k^{n_s-4}$ with $k=|\vec k|$ denotes the initial power spectrum
and $T_l(k)$ is the transfer function.
The transfer function is computed for the cosmological parameters given
in section \ref{Hantzsche-Wendt_Topology} using the full Boltzmann physics
\cite{Ma_Bertschinger_1995,Hu_1995}.
The program takes into account the ordinary and the integrated Sachs-Wolfe
contribution, the Doppler contribution, Silk damping, reionization,
polarisation of photons and neutrinos with standard thermal history.
The code yields the same angular power spectrum as
CAMB\footnote{The software is available at http://camb.info}
for the simply connected space.

For a single realisation of an CMB sky,
the multipole moments are given by
\begin{equation}
\label{Eq:Def_Cl}
C_l \; := \;
\frac 1{2l+1} \sum_{m=-l}^l |a_{lm}|^2
\end{equation}
with $a_{lm}$ obtained from eq.(\ref{Eq:alm}).
The ensemble average simplifies the expression to
\begin{eqnarray}
\label{Eq:Def_Cl_ensemble}
C_l \; = \;
\sum_{\vec k} \frac{T_l^2(k)\,P(k)}{4\pi} \; \Big[ \; 1 
\\ & & \nonumber \hspace{-35pt} + \;
 P_l(\hat k_1\cdot\hat k_2) (-1)^{n_x} \cos(2\pi n_y y) \cos(2\pi n_z z)
\\ & & \nonumber \hspace{-35pt} + \;
 P_l(\hat k_1\cdot\hat k_3) (-1)^{n_y+n_z} \cos(2\pi n_x x) \cos(2\pi n_z z)
\\ & & \nonumber \hspace{-35pt} + \;
 P_l(\hat k_2\cdot\hat k_3) (-1)^{n_x+n_y+n_z} \cos(2\pi n_x x) \cos(2\pi n_y y)
\; \Big]
\end{eqnarray}
where $n_x$, $n_y$, and $n_z$ are determined by $\vec k$ via eq.\,(\ref{Eq:wave_number}).
The Legendre functions are denoted by $P_l(x)$.
The normalisation of the initial power spectrum $P(k)\sim k^{n_s-4}$ is determined by
fitting the ensemble averaged multipole moments (\ref{Eq:Def_Cl_ensemble}) to the
power spectrum from $l=50$ to $l=400$ based on the WMAP seven year data.
This is at sufficiently small angular scales, such that the influence of
the cosmic topology on the large-angle correlations does not alter the
normalisation of $P(k)$,
so that we fit to the standard local physics.

The multipole moments $C_l$ are related to the
full-sky temperature 2-point correlation function
\begin{equation}
\label{Eq:C_theta}
C(\vartheta) \; := \; \left< \delta T(\hat n) \delta T(\hat n')\right>
\hspace{10pt} \hbox{with} \hspace{10pt}
\hat n \cdot \hat n' = \cos\vartheta
\hspace{10pt} ,
\end{equation}
by the equation
\begin{equation}
\label{Eq:C_theta_C_l}
C(\vartheta) \; = \; \sum_l \frac{2l+1}{4\pi} \, C_l \, P_l(\cos\vartheta)
\hspace{10pt} .
\end{equation}
The brackets in equation (\ref{Eq:C_theta}) denote an averaging over
all directions $\hat n$ and $\hat n'$ with an angular separation $\vartheta$.
Since the Hantzsche-Wendt space is not statistically isotropic,
this averaging leads to an information lost compared to the
correlation function $C_{ll'}^{mm'} :=\left< a_{lm} a_{l'm'}^\star \right>$.
The anisotropic correlation leads to further constraints for a
given topology, see e.\,g.\
\cite{Bond_Pogosyan_Souradeep_1998,Inoue_Sugiyama_2003,%
Aurich_Janzer_Lustig_Steiner_2007,Fabre_Prunet_Uzan_2013,%
Aslanyan_Manohar_Yadav_2013,Planck_Topo_2013}.
Thus, we note that the consistency with the statistically isotropic measure
$C(\vartheta)$ is not sufficient, since statistics measuring the
statistically anisotropic properties in the CMB maps could conflict
with the observations.
However, since an analysis of the deviations from statistical isotropy
is computationally too demanding for an inhomogeneous manifold,
we restrict us to the correlation (\ref{Eq:C_theta}).
As already emphasised, this correlation measure is already unusually low
\cite{Copi_Huterer_Schwarz_Starkman_2013a}.


\begin{figure}
\begin{center}
\vspace*{-30pt}
\begin{minipage}{12cm}
\includegraphics[width=12.0cm]{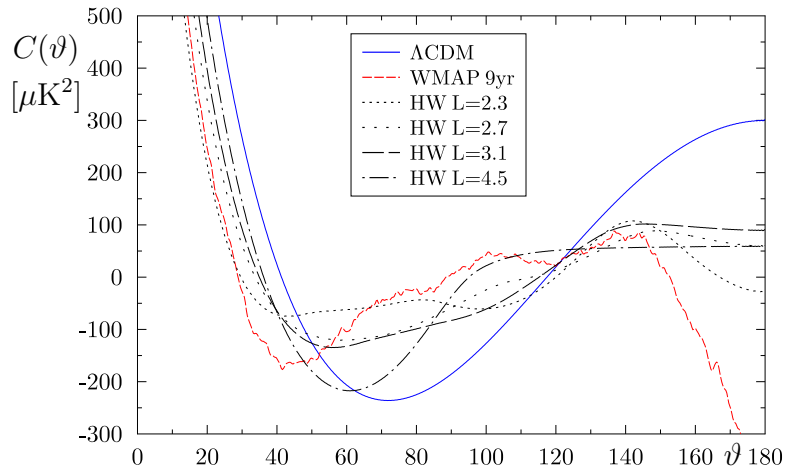}
\end{minipage}
\put(-68,-88){$\vartheta$}
\put(-337,66){$C(\vartheta)$}
\put(-338,48){[$\mu\hbox{K}^2$]}
\end{center}
\vspace*{-30pt}
\caption{\label{Fig:ctheta_ensemble}
The two-point angular correlation function $C(\vartheta)$
is plotted for the standard $\Lambda$CDM model and for
those derived from the WMAP 9yr angular power spectrum $C_l$.
Furthermore, four correlation functions are shown for
the regular Hantzsche-Wendt manifold having a size $L=2.3$, 2.7, 3.1,
and 4.5.
The observer positions $\vec x_o$ with the smallest median of the $S(60^\circ)$
distribution are chosen, see figure \ref{Fig:s60_aus_ensemble_ctheta_positions}.
}
\end{figure}


Figure \ref{Fig:ctheta_ensemble} shows the correlation function $C(\vartheta)$
for the $\Lambda$CDM concordance model as well as those obtained from
the unbinned WMAP 9yr angular power spectrum $C_l$
\cite{Bennett_et_al_2012} using equation (\ref{Eq:C_theta_C_l}).
Furthermore, the correlation $C(\vartheta)$ is plotted for
the Hantzsche-Wendt space using the multipole moments (\ref{Eq:Def_Cl})
in equation (\ref{Eq:C_theta_C_l}) for the length parameters
$L=2.3$, 2.7, 3.1, and 4.5 with $L=L_x=L_y=L_z$.
Here and in the following, we mean with the $\Lambda$CDM concordance model
the model with the $E_{18}$ topology, while both the $\Lambda$CDM model
as well as the Hantzsche-Wendt space are considered for the same
set of cosmological parameters.

\section{Suppression of Large-Scale CMB Correlations}

The COBE mission made the important observation
that the temperature fluctuations $\delta T(\hat n_r)$
are almost uncorrelated above angular scales of $\vartheta=60^\circ$
\cite{Hinshaw_et_al_1996}.
The lack of correlations is unexpected since the $\Lambda$CDM concordance
model predicts much higher correlations.
This strange behaviour persisted every time, new and improved CMB data
were released.
An analysis of the correlations $C(\vartheta)$ concerning various
combinations of recent CMB data sets and diverse masks can be found in
\cite{Copi_Huterer_Schwarz_Starkman_2008,Copi_Huterer_Schwarz_Starkman_2013a}.
The discrepancy between the CMB data and the concordance model
is also illustrated by figure \ref{Fig:ctheta_ensemble}
which emphasises the large values predicted by the $\Lambda$CDM model.
The lack of correlations above $\vartheta=60^\circ$ is conveniently described
by the scalar statistical measure
\begin{equation}
\label{Eq:S_statistic_60}
S(60^\circ) \; := \; \int^{\cos(60^\circ)}_{\cos(180^\circ)}
\hbox{d} \cos\vartheta \; |C(\vartheta)|^2
\end{equation}
which is introduced in \cite{Spergel_et_al_2003}.
A compilation of the values of $S(60^\circ)$
based on various CMB data sets and masks can be found in
\cite{Copi_Huterer_Schwarz_Starkman_2008,Copi_Huterer_Schwarz_Starkman_2013a}.
The latter paper concludes that $C(\vartheta)$ is ``anomalously low
in all relevant maps since the days of the COBE-DMR''. 

Since compact multiply connected spaces possess a lower cut-off in their
wave number spectrum $\{\vec k\,\}$, see for example (\ref{Eq:wave_number}),
they could provide a natural explanation for the lack of correlations.
In the following an analysis is presented for the regular
Hantzsche-Wendt space for which the three lengths $L_x$, $L_y$, and $L_z$
are equal.

\subsection{Full-sky CMB analysis}

In this section we study the full-sky correlations obtained
from numerous simulations of CMB maps based on the Hantzsche-Wendt topology.
Models with the topological length scale $L=L_x=L_y=L_z$ are studied
from $L=1.5$ to $L=6.0$ in steps of $\Delta L=0.1$.
Between $L=6.0$ and $L=9.0$ a step size of $\Delta L=0.5$ is chosen.
As already emphasised, the considered topology is inhomogenous and
requires a CMB analysis for a large set of observer positions $\vec x_o$.
Using the dimensionless coordinates  $(x,y,z)$ with
$\vec x := (x L_x, y L_y, z L_z)$, observer positions on the cube with
$x,y,z \in [0,0.5]$ with a step size of 0.01 are chosen.
This leads to $51^3$ positions $\vec x_o$ for which the CMB is computed
for each value of $L$.
For each position $\vec x_o$ and each value of $L$, a set of 50\,000
simulations is generated according to the multipole spectrum
(\ref{Eq:Def_Cl}) and the $S(60^\circ)$ statistic is calculated.
This allows the determination of the median of the distribution
of $S(60^\circ)$, so that the probability to observe a value larger than the
median or a smaller one is 50\% in either case.
Since the distribution of $S(60^\circ)$ is asymmetrical,
the median is preferred to the mean.
We thus obtain the median of $S(60^\circ)$ for each of the $51^3$ positions
from which the mean value over all observer positions is computed
for fixed $L$.
Furthermore, the maximum and the minimum of the position dependent median
is determined, and the position is stored of that observer
to whom belongs the largest or the smallest median.

\begin{figure}
\begin{center}
\vspace*{-30pt}
\begin{minipage}{12cm}
\includegraphics[width=12.0cm]{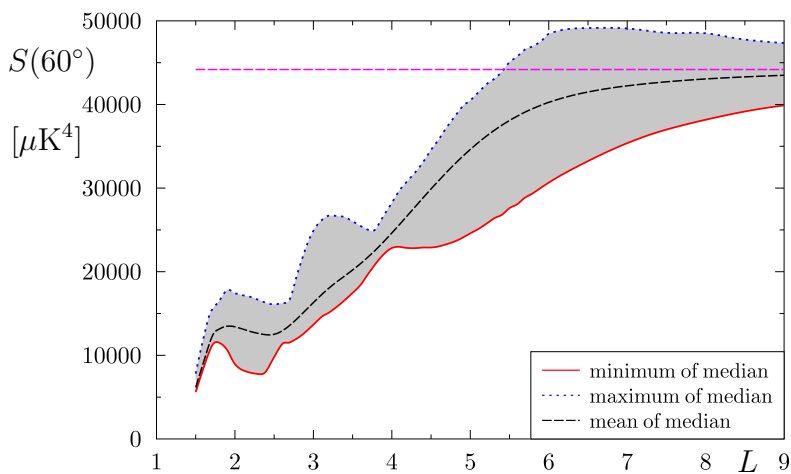}
\end{minipage}
\put(-70,-89){$L$}
\put(-346,63){$S(60^\circ)$}
\put(-345,33){[$\mu\hbox{K}^4$]}
\end{center}
\vspace*{-30pt}
\caption{\label{Fig:s60_aus_50k_simul}
The width of the distribution of the position dependent median
of $S(60^\circ)$ is shown in dependence on the size $L$
of the regular Hantzsche-Wendt manifold.
For each value of $L$, an ensemble of 50\,000 CMB simulations is generated
for each of the $51^3$ observer positions within the manifold,
so that a distribution of $51^3$ median values of $S(60^\circ)$ is obtained
from which the mean, the minimum and the maximum is determined.
The dashed horizontal line at $44182\mu\hbox{K}^4$ belongs to the
median of the $\Lambda$CDM concordance model.
Since there is no position dependence in this case, only one curve
can be plotted.
}
\end{figure}

\begin{figure}
\begin{center}
\vspace*{-15pt}
\begin{minipage}{18cm}
\hspace*{-20pt}\includegraphics[width=10.0cm]{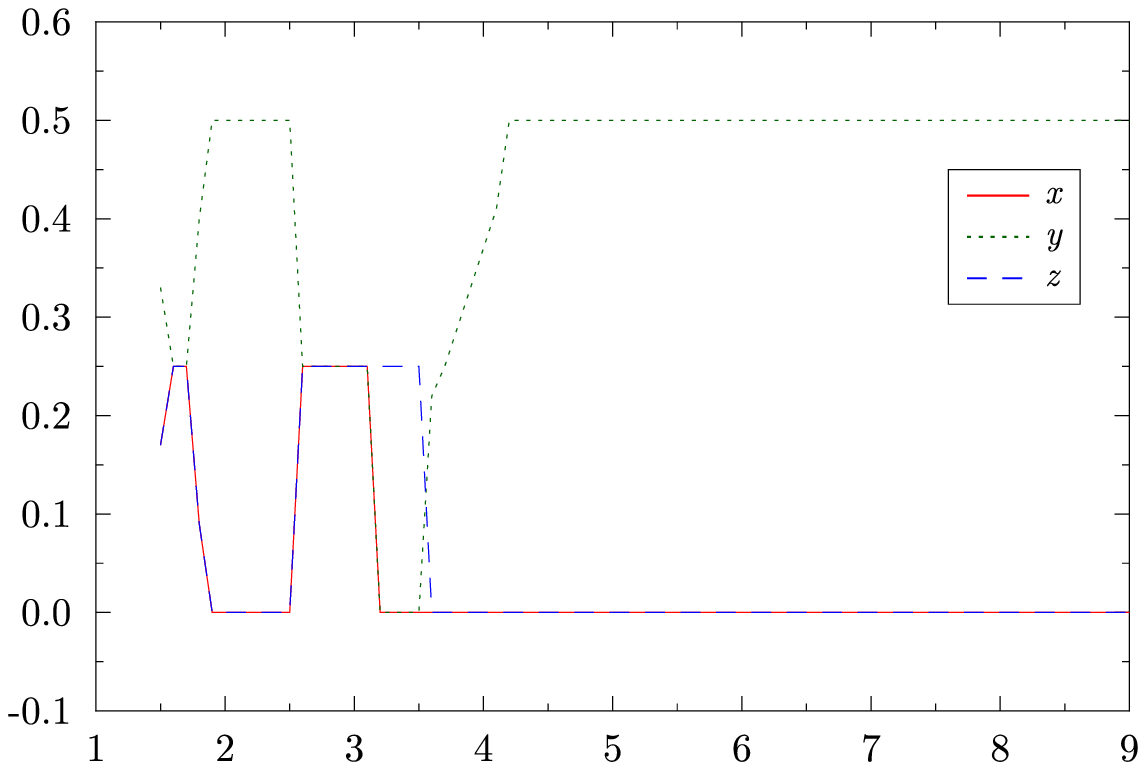}
\hspace*{-40pt}\includegraphics[width=10.0cm]{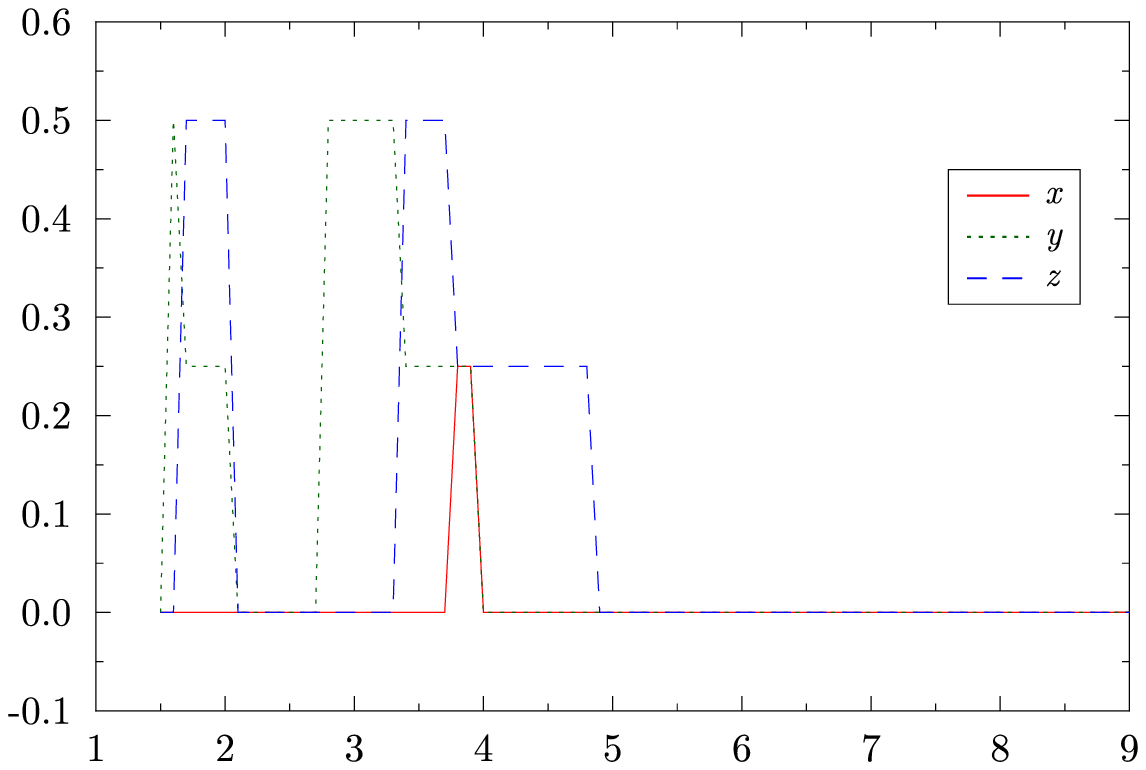}
\end{minipage}
\put(-484,50){$(a)$}
\put(-150,50){$(b)$}
\put(-60,-71){$_L$}
\put(-306,-71){$_L$}
\end{center}
\vspace*{-20pt}
\caption{\label{Fig:s60_aus_ensemble_ctheta_positions}
Panel (a) shows the positions $(x,y,z)$ of that observer
for which the minimum of the median of $S(60^\circ)$ is obtained
as plotted in figure \ref{Fig:s60_aus_50k_simul}.
Panel (b) reveals the corresponding positions for the maximum.
}
\end{figure}

The results are shown in figures \ref{Fig:s60_aus_50k_simul} and
\ref{Fig:s60_aus_ensemble_ctheta_positions}.
Figure \ref{Fig:s60_aus_50k_simul} reveals the suppression of correlations
compared to the $\Lambda$CDM concordance model,
whose $S(60^\circ)$ median is plotted as a horizontal line.
For lengths below $L\simeq 4$, the large-scale correlations of the
Hantzsche-Wendt space are at least a factor of two smaller
than those of the $\Lambda$CDM model.
For the minimum of the median, the suppression is remarkable even
up to length scales of almost $L=5$ where the figure reveals a plateau.
The diameter of the surface of last scattering is
$D_{\hbox{\scriptsize sls}} = 6.605$,
so that the suppression is less pronounced for $L>D_{\hbox{\scriptsize sls}}$
as it is confirmed by figure \ref{Fig:s60_aus_50k_simul}.
At $L=9$ the mean of the median of $S(60^\circ)$ nearly agrees
with the $\Lambda$CDM value.
We showed in section \ref{Hantzsche-Wendt_Topology}
that the Hantzsche-Wendt space with $L>3$ will not be
discovered by searches for matched back-to-back circle pairs.
In the length interval $L=3$ to $L=6$ exists only one matched back-to-back
circle pair with radius $\alpha>25^\circ$ and that only for
very special observer positions $\vec x_o$.
Therefore, there exists a range of sizes of the Hantzsche-Wendt space
having a significant suppression of large-scale correlations without
betraying themselves by back-to-back circle pairs
for general observers.

The positions $(x,y,z)$ of the observers for which the lowest and the
highest values of the median of $S(60^\circ)$ are found,
can be read off from figure \ref{Fig:s60_aus_ensemble_ctheta_positions}.
Whereas panel (a) shows the positions for the smallest values
of the median, panel (b) plots the positions for the maximum.
One observes several switches for these positions.
From $L=2.6$ to $L=3.1$, small medians occur at
$(x,y,z)=\big(\frac 14,\frac 14,\frac 14\big)$.
Around $L=3$ large medians belong to $(x,y,z)=\big(0,\frac 12,0\big)$.
It is interesting to note that exactly the position $\big(0,\frac 12,0\big)$
corresponds to small medians for large spaces $L\geq 4.2$
which persists up to the largest investigated spaces with $L=9$.
This position also leads to small medians between $L=2$ and $L=2.5$.
Since the two positions $(x,y,z)=\big(\frac 14,\frac 14,\frac 14\big)$
and $\big(0,\frac 12,0\big)$ are distinguished,
we will investigate them in more detail below.
The Dirichlet domain for these two cases is shown in
figure \ref{Fig:HW_Dirichlet}, panels (b) and (d).

\begin{figure}
\begin{center}
\vspace*{-30pt}
\begin{minipage}{12cm}
\includegraphics[width=12.0cm]{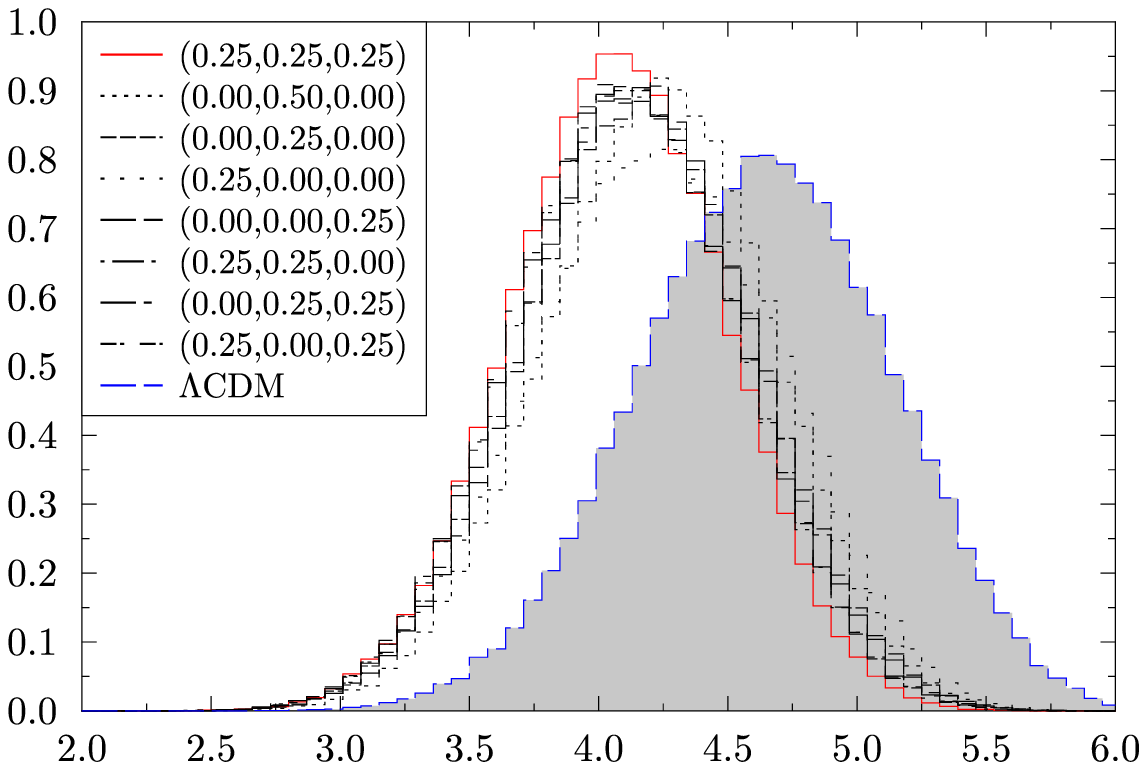}
\put(-34,24){$\log_{10}(S(60^\circ))$}
\put(-80,170){(a)}
\put(-275,80){$L=2.7$}
\end{minipage}
\begin{minipage}{12cm}
\vspace*{-30pt}\includegraphics[width=12.0cm]{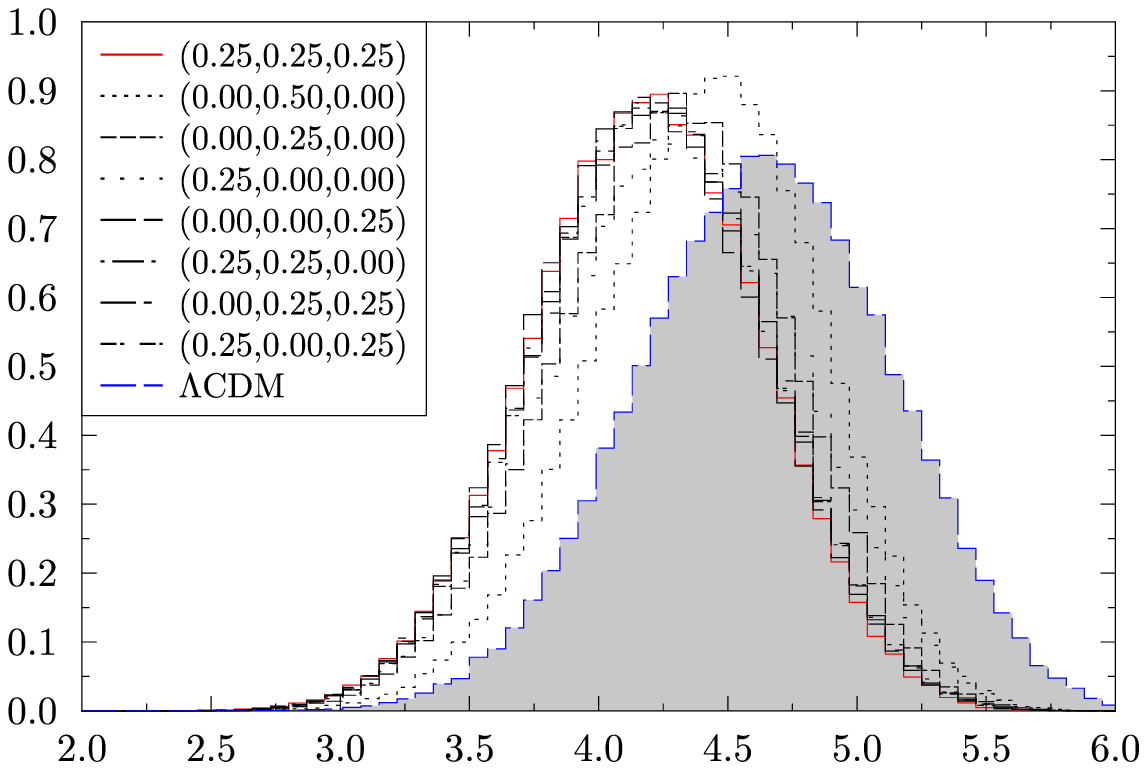}
\put(-34,24){$\log_{10}(S(60^\circ))$}
\put(-80,170){(b)}
\put(-275,80){$L=3.1$}
\end{minipage}
\begin{minipage}{12cm}
\vspace*{-30pt}\includegraphics[width=12.0cm]{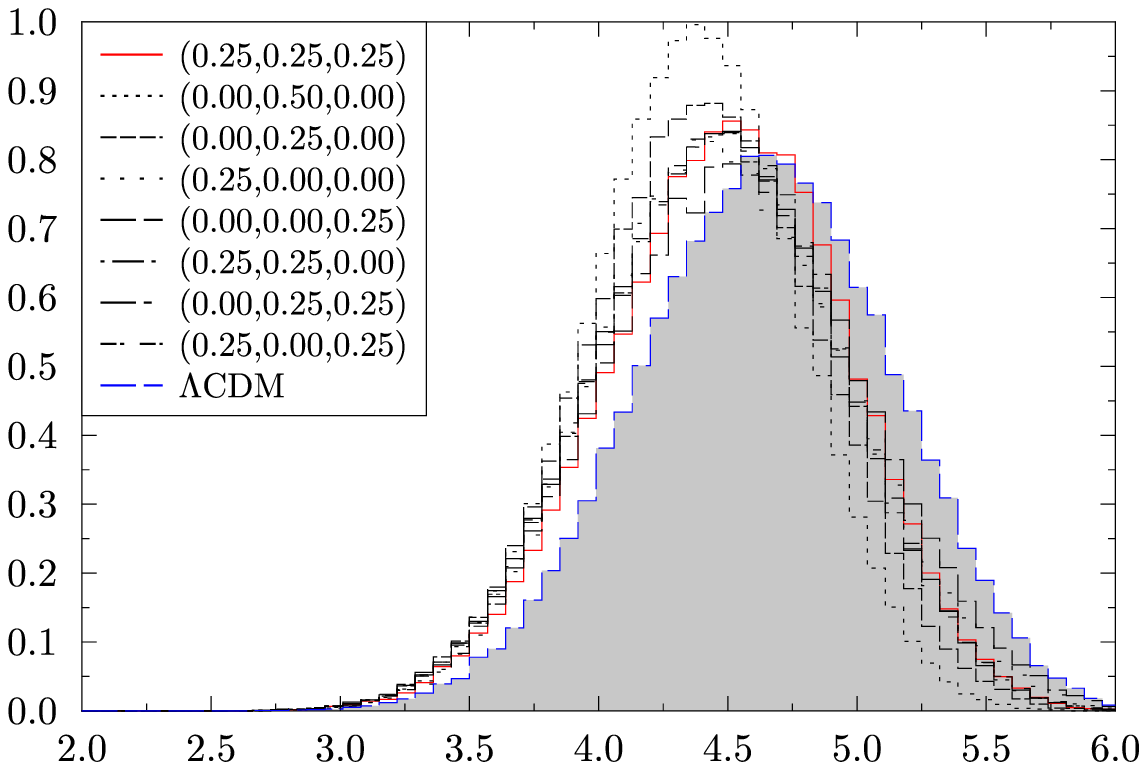}
\put(-34,24){$\log_{10}(S(60^\circ))$}
\put(-80,170){(c)}
\put(-275,80){$L=4.5$}
\end{minipage}
\end{center}
\vspace*{-25pt}\caption{\label{Fig:s60_histo_nomask}
The distribution of $\log_{10}(S(60^\circ))$ is shown for three different
topological lengths $L$ in panels (a) to (c).
The histograms are computed from 100\,000 simulations for various
positions $(x,y,z)$ of the observer as stated in the legend.
In addition, all panels show the corresponding distribution of
the $\Lambda$CDM model.
}
\end{figure}

Figure \ref{Fig:s60_aus_50k_simul} shows the width of the
distribution of the medians of $S(60^\circ)$,
and thus does not reveal the distribution of the medians for
a fixed observer position $\vec x_o$.
This information is provided by figure \ref{Fig:s60_histo_nomask}
which displays the distribution for several observer positions $\vec x_o$.
Since the distributions are very asymmetrical having a pronounced
tail towards large values,
the histograms are plotted using a logarithmic scale.
With such a scaling the distributions virtually look almost like
Gaussians, but they are not, of course.
For the three topological scales $L=2.7$, $L=3.1$, and $L=4.5$,
the histograms belonging to 8 different positions are shown
in comparison to the distribution of the $\Lambda$CDM model
which is shaded in grey.
All histograms are based on 100\,000 simulations and reveal
the position dependence of the distributions.
The histograms show the degree of how much the distributions of
the Hantzsche-Wendt space are shifted towards smaller medians of $S(60^\circ)$
compared to the $\Lambda$CDM model.
Note that due to the logarithmic scale, the difference between the
Hantzsche-Wendt space and the infinite space is more pronounced
than figure \ref{Fig:s60_histo_nomask} suggests.
The switch of the positions belonging to extreme medians described
in connection with figure \ref{Fig:s60_aus_ensemble_ctheta_positions}
can also be seen in figure \ref{Fig:s60_histo_nomask}.
Panels (a) and (b), belonging to $L=2.7$ and $L=3.1$, respectively, show
that the histogram belonging to $(x,y,z)=\big(\frac 14,\frac 14,\frac 14\big)$
is shifted towards small medians,
while the histogram belonging to $\big(0,\frac 12,0\big)$ is shifted
towards the $\Lambda$CDM histogram.
Panel (c), belonging to $L=4.5$, shows the reverse behaviour so that
the distribution due to $\big(0,\frac 12,0\big)$ has smaller medians
as well as the most pronounced peak.

For $L=2.7$ there are 3 back-to-back circle pairs with
a radius of $35^\circ$, and for the position $\vec x_o=\big(\frac 14,0,0\big)$
lying on the axis of rotation of the first generator in (\ref{Eq:Def_Gamma})
exists an additional one with $66^\circ$
(compare figure \ref{Fig:cits_back_to_back}).
The larger fundamental domain with $L=3.1$ does not possess
back-to-back circle pairs with radii above $25^\circ$
except for the position $\vec x_o=\big(\frac 14,0,0\big)$
where one occurs with radius $62^\circ$.
The same applies for the space with $L=4.5$
except that for the special position the radius is reduced to $47^\circ$.

Since the CMB observations point to an exceedingly small value
of $S(60^\circ)$,
the most important detail in the histograms lies in the left tail.
The paper \cite{Copi_Huterer_Schwarz_Starkman_2013a} discusses only
the lack of large-angle correlations in the cut sky,
but the values listed in their tables 1 and 2 are always below
$1900\mu\hbox{K}^4$ for different combinations of CMB measurements
and masks.
Although these $S(60^\circ)$ values refer to the cut sky,
we nevertheless compute the corresponding values for our full-sky
simulations.
The cut-sky maps are discussed in the next section.

\begin{figure}
\begin{center}
\vspace*{-30pt}
\begin{minipage}{12cm}
\includegraphics[width=12.0cm]{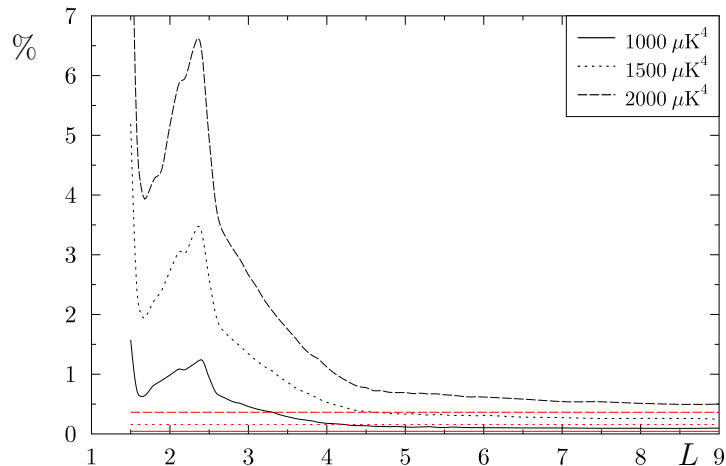}
\end{minipage}
\put(-68,-88){$L$}
\put(-320,66){$\%$}
\end{center}
\vspace*{-30pt}
\caption{\label{Fig:s60_schwelle}
The probability given in percent $\%$ is plotted for obtaining
a model with an value of $S(60^\circ)$ below the threshold
$1000\mu K^4$, $1500\mu K^4$, and $2000\mu K^4$, respectively.
The probabilities are obtained from 100\,000 simulations
and the value is taken for that observer position
for which the largest probability occurs.
The straight horizontal lines give the corresponding
probabilities of the $\Lambda$CDM model.
}
\end{figure}

To calculate the likelihood that a given model has a correlation of
$S(60^\circ)$ below a given threshold $S_{\hbox{\scriptsize threshold}}$,
a set of 100\,000 CMB simulations is generated and the number of simulations
having $S(60^\circ)<S_{\hbox{\scriptsize threshold}}$ gives the probability
$p(S_{\hbox{\scriptsize threshold}})$.
For the $\Lambda$CDM model, one gets the probabilities
$p(1000\mu\hbox{K}^4)=0.044\%$, $p(1500\mu\hbox{K}^4)=0.158\%$, and
$p(2000\mu\hbox{K}^4)=0.363\%$.
For the Hantzsche-Wendt space we compute the probabilities for these
thresholds for the observer position $\vec x_o$ belonging to
the minimum of median of the $S(60^\circ)$ distribution as shown in
figure \ref{Fig:s60_aus_ensemble_ctheta_positions}.
Throughout the considered $L$ interval the Hantzsche-Wendt probabilities
are higher than the corresponding ones of the $\Lambda$CDM model,
as can be seen in figure \ref{Fig:s60_schwelle}.
As expected for $L<D_{\hbox{\scriptsize sls}}$ the occurrences of small medians
are much more likely for the Hantzsche-Wendt space
than for the $\Lambda$CDM model.
A pronounced peak probability occurs around $L\simeq 2.4$,
see figure \ref{Fig:s60_schwelle}.
Let us now turn to a comparison which takes a mask into account.

\subsection{Analysis with WMAP mask kq75 9yr}

In this section we analyse the reduction of large-scale correlations of the
Hantzsche-Wendt topology in the presence of a mask.
As pointed out in \cite{Copi_Huterer_Schwarz_Starkman_2013a}
the correlations on the cut sky are already unusually low,
so that it is not compelling to delve into a full-sky reconstruction
which will burden the analysis with unknown biases.
Thus we will now focus on the cut-sky correlations
which will be derived analogous to the observational data
in order to provide an internally consistent comparison of the
large-scale correlations.

The correlations of the Hantzsche-Wendt space are obtained according
the following pipeline.
At first the expansion coefficients of the temperature anisotropy
(\ref{Eq:CMB_expansion}) are computed using
\begin{equation}
\label{Eq:HW_alm}
a_{lm} \; = \;
\sum_{\vec k} T_l(k) \, \sqrt{P(k)} \, \xi_{lm}^{\vec k}(\vec x_o)
\hspace{10pt} ,
\end{equation}
where $\xi_{lm}^{\vec k}(\vec x_o)$ is given by equation (\ref{Eq:a_lm}).
Since the generators (\ref{Eq:Def_Gamma}) single out the axes of the
coordinate system, the CMB radiation is statistically anisotropic.
Therefore, a random rotation is applied to the coefficients $a_{lm}$
which is done using the Wigner $D$ matrices $D_{m_1m_2}^l(\alpha,\beta,\gamma)$,
where $\alpha,\beta,\gamma$ denote the Euler angles of the rotation.
Choosing the rotation angles $\alpha,\beta,\gamma$ uniformly distributed
over their definition interval,
one gets simulated sky maps with a random orientation of the fundamental cell.
In the next step the monopole and dipole is removed.
Thereafter a full-sky map is generated and the kq75 9year mask of the
WMAP team \cite{Bennett_et_al_2012} is applied.
From this cut-sky map the coefficients $\widetilde{a}_{lm}$ are computed
which in turn lead to the pseudo-$C_l$
\begin{equation}
\label{Eq:Cl_pseudo}
\widetilde{C}_l \; := \;
\frac 1{2l+1} \sum_{m=-l}^l |\widetilde{a}_{lm}|^2
\hspace{10pt} .
\end{equation}
This allows the computation of the pseudo-correlation function
\begin{equation}
\label{Eq:C_theta_pseudo}
\widetilde{C}(\vartheta) \; = \; \frac 1{A(\vartheta)} \,
\sum_{l=0}^{l_{\hbox{\scriptsize max}}} \, \frac{2l+1}{4\pi} \, \widetilde{C}_l \, P_l(\cos\vartheta)
\hspace{10pt} ,
\end{equation}
where
\begin{equation}
\label{Eq:A_theta}
A(\vartheta) \; := \;
\sum_{l=0}^{l_{\hbox{\scriptsize max}}} \, \frac{2l+1}{4\pi} \, A_l \, P_l(\cos\vartheta)
\hspace{10pt} \hbox{ with } \hspace{10pt}
A_l := \frac 1{2l+1} \, \sum_{m=-l}^l |w_{lm}|^2
\end{equation}
takes the correct normalisation into account.
The spherical expansion coefficients of the mask are denoted by $w_{lm}$.
Thereafter the final step is the calculation of $S(60^\circ)$
from $\widetilde{C}(\vartheta)$.

\begin{figure}
\begin{center}
\vspace*{-20pt}
\begin{minipage}{18cm}
\hspace*{-50pt}\includegraphics[width=10.0cm]{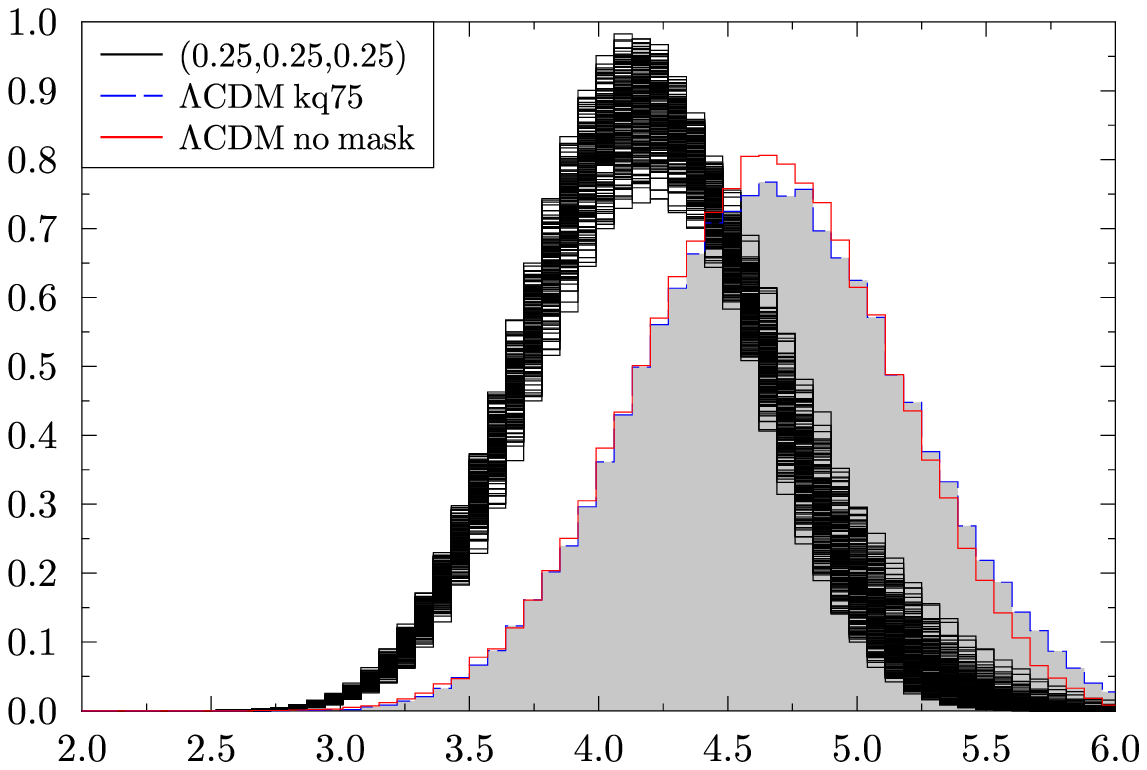}
\hspace*{-50pt}\includegraphics[width=10.0cm]{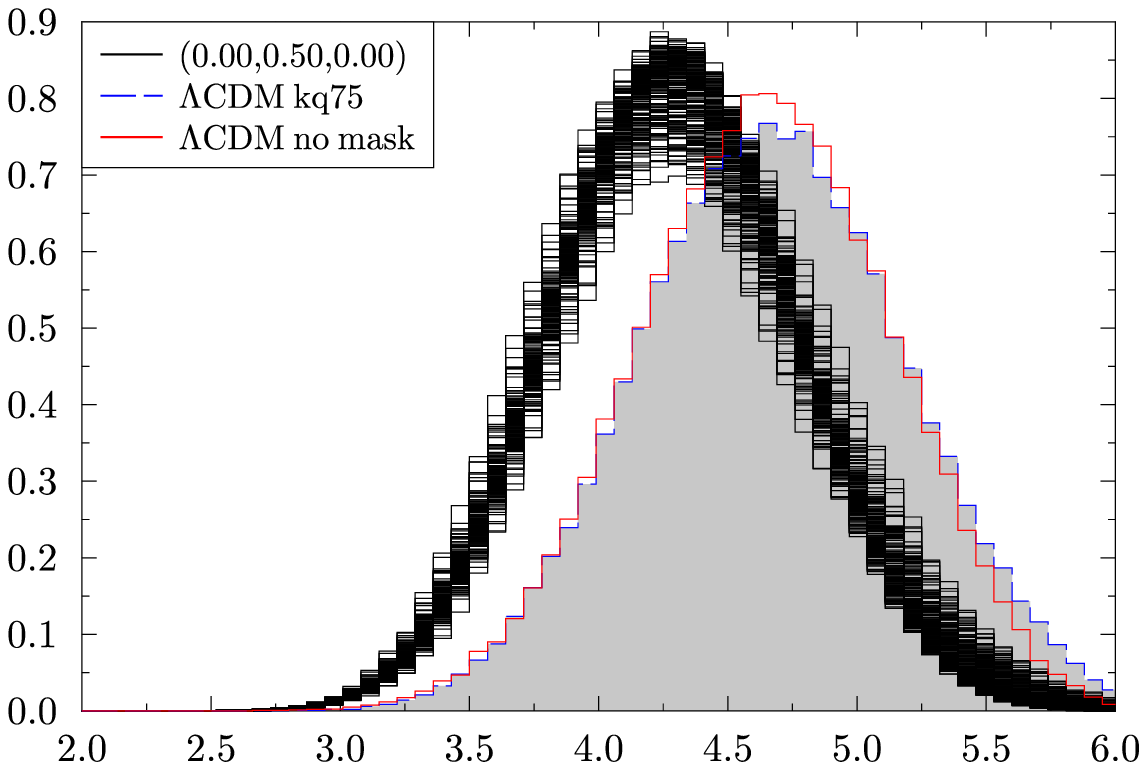}
\put(-34,20){\small{$\log_{10}(S(60^\circ))$}}
\put(-320,140){(a)}
\put(-70,140){(b)}
\put(-468,100){\small{$L=2.7$}}
\put(-230,100){\small{$L=2.7$}}
\end{minipage}
\begin{minipage}{18cm}
\vspace*{-25pt}\hspace*{-50pt}\includegraphics[width=10.0cm]{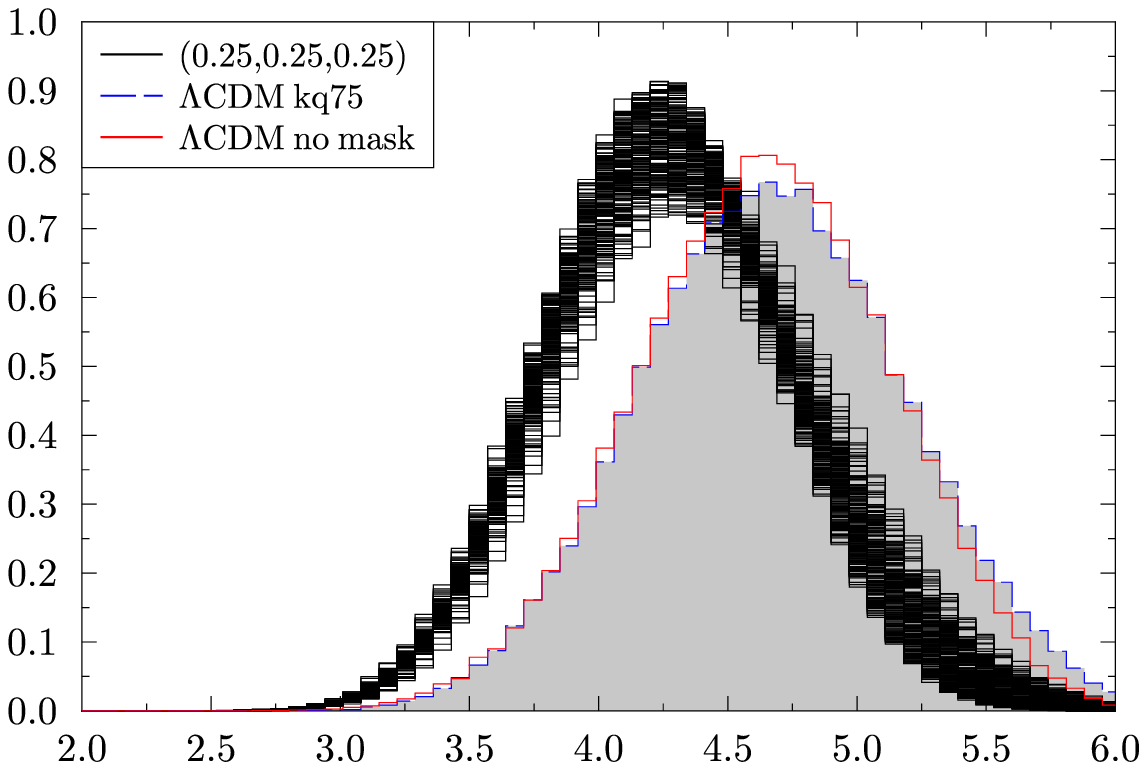}
\hspace*{-50pt}\includegraphics[width=10.0cm]{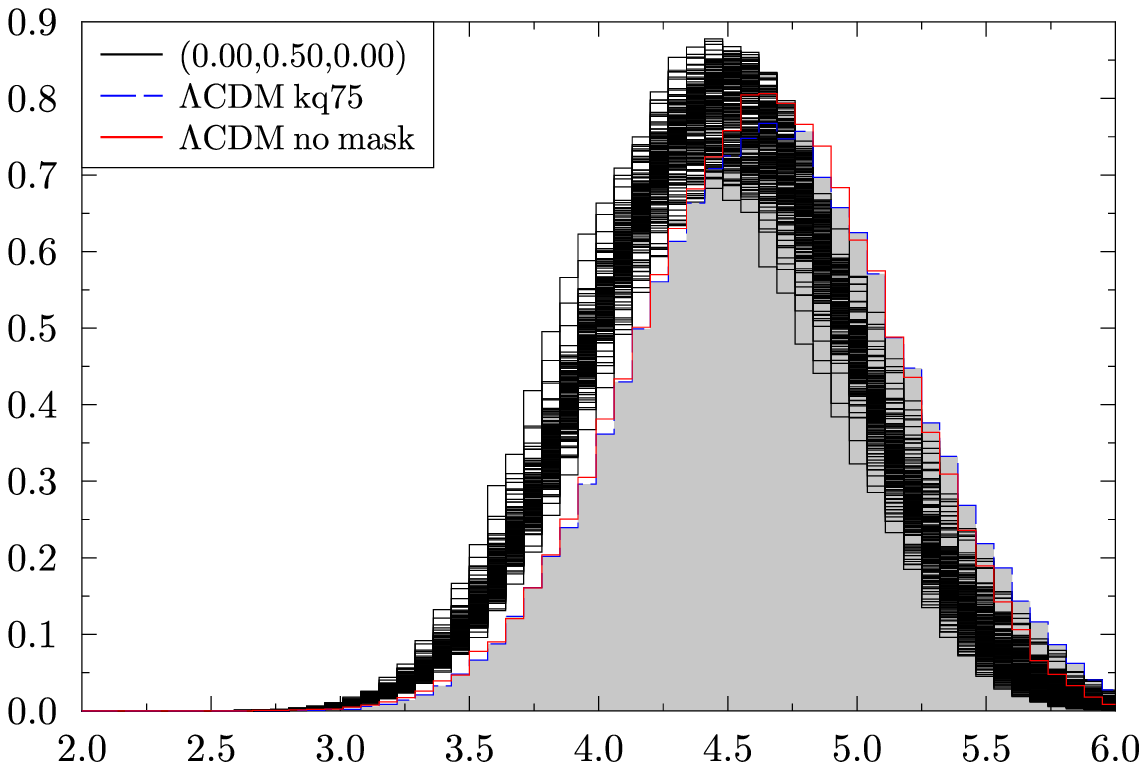}
\put(-34,20){\small{$\log_{10}(S(60^\circ))$}}
\put(-320,140){(c)}
\put(-70,140){(d)}
\put(-468,100){\small{$L=3.1$}}
\put(-230,100){\small{$L=3.1$}}
\end{minipage}
\begin{minipage}{18cm}
\vspace*{-25pt}\hspace*{-50pt}\includegraphics[width=10.0cm]{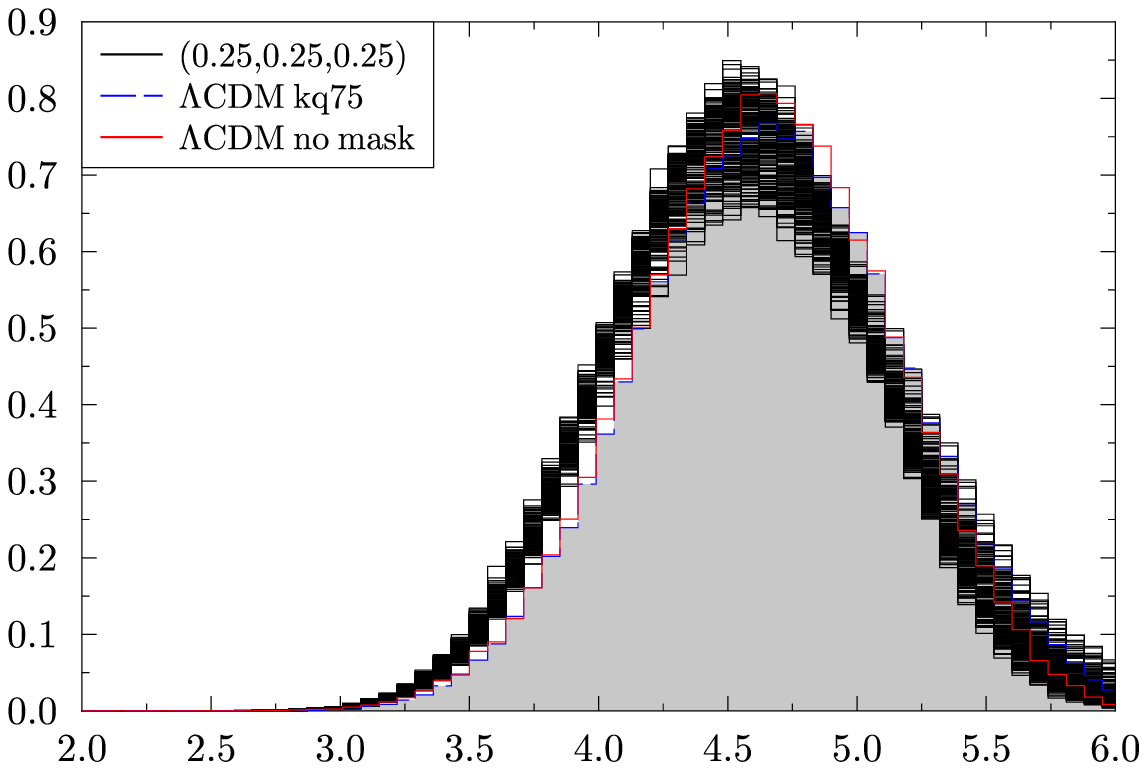}
\hspace*{-50pt}\includegraphics[width=10.0cm]{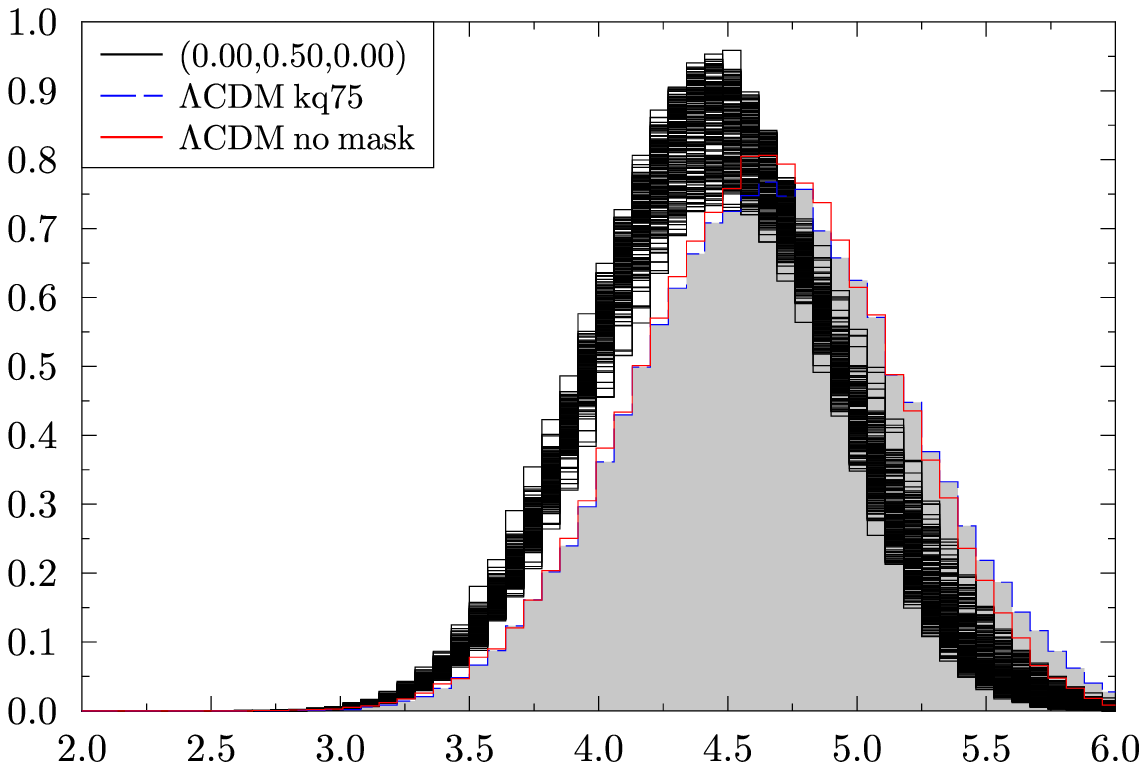}
\put(-34,20){\small{$\log_{10}(S(60^\circ))$}}
\put(-320,140){(e)}
\put(-70,140){(f)}
\put(-468,100){\small{$L=4.5$}}
\put(-230,100){\small{$L=4.5$}}
\end{minipage}
\end{center}
\vspace*{-20pt}\caption{\label{Fig:s60_histo_kq75}
The distribution of $\log_{10}(S(60^\circ))$ is shown for three different
topological lengths $L$ in panels (a) to (f).
In contrast to figure \ref{Fig:s60_histo_nomask}, the distributions are
computed for 200 random orientations of simulated sky maps
relative to the kq75 mask.
The histograms are based on 100\,000 simulations for the observer
position $(x,y,z)$ stated in the legend.
In addition, all panels show the corresponding distribution of
the $\Lambda$CDM model using the kq75 mask.
To demonstrate the modest influence of the mask in the case of the
$\Lambda$CDM simulations we also show its distribution without a mask. 
}
\end{figure}

We investigate the distribution of $S(60^\circ)$ for the
regular Hantzsche-Wendt space with $L=2.7$, $3.1$, and $4.5$.
The two special points $(x,y,z)=\big(\frac 14,\frac 14,\frac 14\big)$
and $\big(0,\frac 12,0\big)$ are selected for the observer positions.
Then 200 different sets of Euler angles $\alpha,\beta,\gamma$ are
randomly generated.
For each of the 200 orientations of the fundamental cell,
we compute the distributions of $S(60^\circ)$ from 100\,000 simulations
as described above.
The results are shown in figure \ref{Fig:s60_histo_kq75}
where an overlay of the 200 histograms is plotted for each case of
$L$ and $\vec x_o$.
In addition, the distribution for the $\Lambda$CDM model is displayed as
the histogram shaded in grey.
The application of the kq75 9yr mask modifies the $\Lambda$CDM histogram
only slightly, as can be inferred from figure \ref{Fig:s60_histo_kq75}.
Furthermore, the figure reveals the anisotropic behaviour of the
cut-sky maps of the Hantzsche-Wendt space,
since the histograms differ only in the orientation of the CMB maps
determined by $\alpha,\beta,\gamma$.
One sees that the histograms are significantly shifted towards smaller
values of $S(60^\circ)$ for $L=2.7$ and $L=3.1$ compared to the
$\Lambda$CDM case.

Although the Hantzsche-Wendt topology possesses significantly smaller
large-angle correlations than the $\Lambda$CDM model,
the most interesting point lies in the tail of the distribution towards
very small values of $S(60^\circ)<2000\mu\hbox{K}^4$.
To discuss that issue we compute the probabilities
$p(S_{\hbox{\scriptsize threshold}})$ from the 200 histograms
for the six cases shown in figure \ref{Fig:s60_histo_kq75}.
The table \ref{Tab:threshold_with_mask} gives the result together
with the probabilities belonging to the $\Lambda$CDM case with
and without the application of a mask.
One observes that, as expected, the probabilities
$p(S_{\hbox{\scriptsize threshold}})$ are always enhanced with respect to
the $\Lambda$CDM model.

\begin{table}[!htbp]
\centering
\begin{tabular}{|c|c|c|c|c|}
\hline
model & position & $p(1000\mu\hbox{K}^4)$ & $p(1500\mu\hbox{K}^4)$  &
$p(2000\mu\hbox{K}^4)$\\
 \hline
HW $L=2.7$ & (0.25, 0.25, 0.25) & 0.235\% &  0.891\% & 1.984\% \\
HW $L=2.7$ & (0.00, 0.50, 0.00) & 0.174\% &  0.665\% & 1.492\% \\ 
 \hline
HW $L=3.1$ & (0.25, 0.25, 0.25) & 0.164\% &  0.642\% & 1.465\% \\
HW $L=3.1$ & (0.00, 0.50, 0.00) & 0.079\% &  0.309\% & 0.719\% \\ 
 \hline
HW $L=4.5$ & (0.25, 0.25, 0.25) & 0.046\% &  0.196\% & 0.481\% \\ 
HW $L=4.5$ & (0.00, 0.50, 0.00) & 0.049\% &  0.211\% & 0.525\% \\ 
 \hline
HW $L=9.0$ & (0.00, 0.50, 0.00) & 0.030\% &  0.131\% & 0.324\% \\ 
 \hline
3-torus $L=4.0$   & independent & 0.118\% &  0.511\% & 1.231\% \\ 
 \hline
$\Lambda$CDM kq75 9yr & independent & 0.028\% & 0.103\% & 0.260\% \\
 \hline
$\Lambda$CDM no mask & independent & 0.044\% & 0.158\% & 0.363\% \\
 \hline
\end{tabular}
\caption{\label{Tab:threshold_with_mask}
The probabilities $p(1000\mu\hbox{K}^4)$, $p(1500\mu\hbox{K}^4)$, and
$p(2000\mu\hbox{K}^4)$ are computed for CMB maps
which are masked using a random orientation relative to the kq75 mask.
The distributions of $S(60^\circ)$ belonging to the first six models
are shown in figure \ref{Fig:s60_histo_kq75}.
Only the last row of the table refers to probabilities computed
without applying a mask.
}
\end{table}

\begin{figure}
\begin{center}
\vspace*{-30pt}
\begin{minipage}{12cm}
\includegraphics[width=12.0cm]{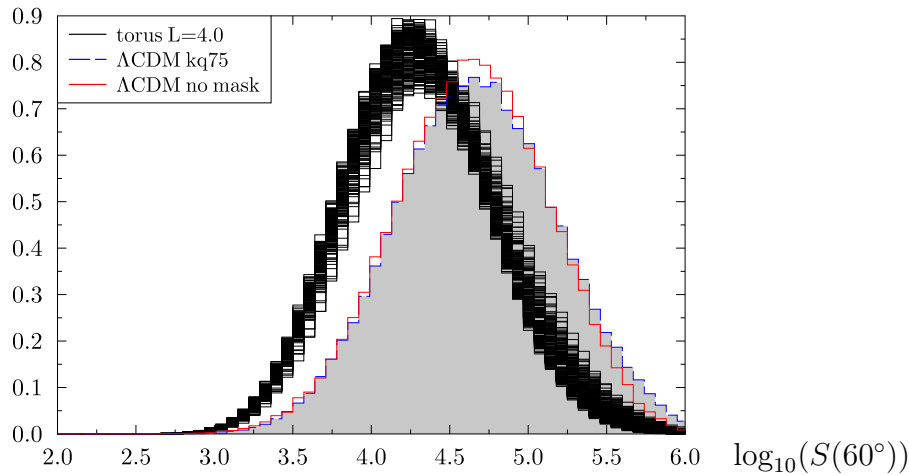}
\end{minipage}
\put(-34,-88){$\log_{10}(S(60^\circ))$}
\end{center}
\vspace*{-30pt}
\caption{\label{Fig:s60_histo_kq75_torus}
The distribution of $\log_{10}(S(60^\circ))$ is shown for the homogeneous
cubic 3-torus topology with $L=4.0$.
The kq75 9yr mask is applied to the torus simulations
similar to figure \ref{Fig:s60_histo_kq75}.
}
\end{figure}

Since the 3-torus topology is well studied, it is interesting to compare it
with the Hantzsche-Wendt topology.
Because the volume of the regular Hantzsche-Wendt space is $2\,L^3$,
i.\,e.\ twice that of the cubic 3-torus having a volume of $L^3$,
the 3-torus length $L=4$ corresponds to the regular Hantzsche-Wendt space
with $L=4/2^{1/3}\simeq 3.17$ by volume.
So we compare the cubic 3-torus with $L=4$ with the Hantzsche-Wendt manifold
with $L=3.1$ in the following.
Although the 3-torus topology is homogeneous,
its CMB radiation is not statistically isotropic.
The same procedure that is applied to cut-sky simulations in the case
of the Hantzsche-Wendt space is also applied for the 3-torus topology.
Their CMB maps were rotated by randomly chosen Euler angles,
and the kq75 mask is used.
Figure \ref{Fig:s60_histo_kq75_torus} shows the histograms for
the cubic 3-torus with $L=4$ analogous to figure \ref{Fig:s60_histo_kq75}
demonstrating the degree of anisotropy.
The probabilities $p(S_{\hbox{\scriptsize threshold}})$ of this cubic 3-torus
topology are also given in table \ref{Tab:threshold_with_mask}.
Since the 3-torus topology is homogeneous, it does not possess a
position dependence, and a single row in the table suffices.
Comparing the torus probabilities with the Hantzsche-Wendt space with
$L=3.1$, one observes that they lie between the probabilities
of the Hantzsche-Wendt observer positions belonging to the
highest and the lowest probability, respectively.
So one concludes that the suppression of large-scale correlations of
the 3-torus is comparable to the Hantzsche-Wendt topology.
However, the 3-torus has at that size several matched back-to-back
circle pairs which contrasts to the Hantzsche-Wendt case as discussed
in section \ref{Hantzsche-Wendt_Topology}.

\section{Summary}

The observed low amplitudes of large-angle temperature correlations
$C(\vartheta)$ could be explained naturally by multiply connected spaces
if their sizes fits well within the surface of last scattering
from which the CMB radiation originates.
Because of missing convincing hints for matched circle pairs in the CMB sky,
the explanation for the suppression of correlations as a consequence
of a non-trivial cosmic topology is currently somewhat disfavoured.
We thus devote this paper to the Hantzsche-Wendt manifold
which is a compact and orientable manifold and lives in the
flat three-dimensional Euclidean space $\mathbb{E}^3$.

It is shown that the regular Hantzsche-Wendt space with lengths
$L\gtrsim 3$ in units of the Hubble length $L_{\hbox{\scriptsize H}}$
possesses only for carefully selected observer positions a single
matched back-to-back circle pair.
For general observers there will be none.
There are non-back-to-back circle pairs, but they are much harder to detect
due to the enhanced background of spurious signals.
It is shown that Hantzsche-Wendt spaces with $L\simeq 3$ possess
large-angle correlations which are reduced around a factor of two or more
in comparison to the $\Lambda$CDM concordance model.
Furthermore, the amplitude of the correlations in the Hantzsche-Wendt topology
is comparable to that of the 3-torus topology,
if spaces of the same volume are considered.
However, in contrast to the Hantzsche-Wendt space, the 3-torus has
matched back-to-back circle pairs at that size.

So we conclude that a Hantzsche-Wendt space around $L\simeq 3$ is a very
interesting topology for the spatial structure of our Universe
which might escape the detection by searches after matched circle pairs
and nevertheless has low large-angle temperature correlations.


\section*{Acknowledgements}

The WMAP data from the LAMBDA website (lambda.gsfc.nasa.gov)
were used in this work.


\section*{References}

\bibliography{../bib_astro}

\bibliographystyle{h-physrev5}

\end{document}